\documentclass{aa}  
\usepackage{graphicx}
\usepackage{txfonts}
\usepackage{lipsum}
\usepackage{subcaption}       
\usepackage{lscape}             
\usepackage{placeins}                                     
\usepackage[colorlinks=true, linkcolor=blue, citecolor=blue]{hyperref}
\usepackage{etoolbox}

\begin{document}
   \title{Sulfur photochemistry traces mantle redox states of rocky planets}

   \author{Ioannis Panagiotou\inst{1}\fnmsep \thanks{Corresponding author: \email{ioannisesc@gmail.com}}
        \and Tim Lichtenberg\inst{1}
        \and Shang-Min Tsai \inst{2}
        \and Harrison Nicholls\inst{3}
        }

   \institute{Kapteyn Astronomical Institute, University of Groningen, PO Box
800, 9700 AV Groningen, The Netherlands
    \and Institute of Astronomy \& Astrophysics, Academia Sinica (ASIAA), Taipei 10617, Taiwan
   \and Institute of Astronomy, University of Cambridge, Cambridge CB3 0HA, United Kingdom}

   \date{Received 15 July 2026 / Accepted 12 November 2026}

\abstract
    {Volatile outgassing from planetary interiors controls the composition of rocky exoplanets' secondary atmospheres. However, observations indicate that disequilibrium processes can strongly alter the chemical structure of hot Jupiters. How outgassing and photochemistry jointly determine the resulting atmospheric composition remains an open question. Sulfur-bearing species are promising tracers of interior-atmosphere coupling because their atmospheric abundances are sensitive to both mantle redox state and stellar irradiation.}
    {We investigate how photochemistry modifies outgassed atmospheres of diverse Earth-sized worlds during the magma ocean phase and assess how sulfur traces the interior oxidation state.}
    {The \texttt{PROTEUS} planetary interior-atmosphere evolution modelling framework is coupled to two chemical models, \texttt{FastChem} and \texttt{VULCAN}. We ran a grid of planetary evolution simulations spanning diverse mantle redox states, instellation fluxes, and host-star spectra. For each case, we compared atmospheric compositions with three post-processing calculations: thermochemical equilibrium, with disequilibrium vertical transport and with vertical transport plus photochemistry.}
    {The bulk atmospheric composition remains controlled by the mantle redox state and the history of outgassing. Reduced mantles produce atmospheres rich in $\mathrm{H_2}$ and $\mathrm{CH_4}$ and oxidised mantles are dominated by $\mathrm{CO_2}$ and $\mathrm{H_2O}$. Photochemistry affects the low-pressure upper atmosphere, strongly depleting neutral volatiles and enhancing radicals. $\mathrm{SO_2}$ is strongly enhanced at intermediate-to-oxidised redox states. Synthetic emission spectra show that photochemical $\mathrm{SO_2}$ can generate absorption features at $4\,\mu$m and at $7.3\,/\,8.7\,\mu$m, reaching $\sim60\,\rm{ppm}$ and $\sim100\,\rm{ppm}$, respectively, before sequentially returning to the outgassed signatures of $\sim30\,\rm{ppm}$ and $\sim50\,\rm{ppm}$ for the oxidised mantle redox state. These signatures are potentially detectable with JWST, motivating targeted observational campaigns.}
    {Sulfur species link mantle redox state, volcanic outgassing, and atmospheric disequilibrium, acting as promising probes of exoplanets' interiors for spectroscopic observations.}


   \maketitle
   \nolinenumbers

\section{Introduction}\label{sec: Introduction}

The number of confirmed exoplanets has risen to more than 6000, revealing a diversity in composition that surpasses that of the Solar System \citep{Winn_2015}. Transit surveys highlight two major populations of rocky planets, super-Earths and sub-Neptunes, with the population deficit between them known as the radius valley \citep{Fulton_2017}. The radius and orbital period occupied by the radius valley hint at sensitive formation and evolution mechanisms \citep{Venturini_2020}, incorporating gas accretion from the stellar nebula, contraction due to cooling \citep{rogers_on_2025, Nicholls_2026}, atmospheric escape processes, orbital migration \citep{shin_super_2026}, and the retention of volatile species collectively affecting the evolution of super-Earths and sub-Neptunes \citep{Owen_2017, Bean_2021, Tang_2024}. Overall, rocky planets exhibit a highly diverse range of interiors and atmospheres due to the many mechanisms driving their evolution \citep{Wordsworth_2022,Lichtenberg2025TrGeo}.

A non-negligible subset of the observed $1$--$10\,M_\oplus$, $0.75$--$3.0\,R_\oplus$ planet population lies in the strongly irradiated lava or rock-vapour regime \citep{Sanchis-Ojeda_2014, Lichtenberg_2025}, suggesting the existence of a surface magma ocean \citep{Kite_2016}. In fact, \citet{Calder_2026} suggests that most sub-Neptunes described by the gas dwarf scenario (i.e. an $\mathrm{MgSiO_3}$ interior surrounded by a thick H/He envelope) remain in the magma ocean phase over their gigayear evolution. Strong irradiation and high temperatures do not necessarily imply a bare-rock surface, since volatile species can be dissolved in the magma ocean and subsequently be outgassed, forming a secondary atmosphere, even under the influence of strong atmospheric escape. The presence of a volatile envelope around low-mass exoplanets has been tentatively suggested for the short-period super-Earths K2--141\,b \citep{Zieba_2022}, 55 Cnc e \citep{Hu_2024}, TOI--431\,b \citep{Monaghan2025AJ}, TOI--561\,b \citep{Teske_2025}, and HD~3167\,b \citep{Coy_2026}. 

To dissipate the leftover internal heat from their formation, all rocky planets are expected to undergo a magma ocean phase in their evolution \citep{Tonks_1993, Elkins-Tanton_2012}. Previous work by \citet{Hirschmann_2012} highlighted the importance of internal redox reactions in shaping the mantle composition and secondary atmospheres of rocky planets produced by magma-ocean evolution and volcanic outgassing. In these systems, it is assumed that the composition of the atmosphere is closely linked to the speciation of volatiles released from the mantle, directly coupling interior redox evolution to atmospheric chemistry \citep{Hirschmann_2012, Schaefer_2017}. The total atmospheric inventory is further influenced by the solubility of volatile species in the magma and by the partitioning between silicate melt and metal, which can retain elements such as carbon in the planetary interior \citep{Hirschmann_2012}. \citet{Lichtenberg_2021} identify two distinct evolutionary pathways governed by magma ocean dynamics and the behaviour of metallic iron droplets. In the first pathway, inefficient magma ocean circulation allows iron droplets to segregate and rain out into the deep mantle, promoting mantle oxidation and the outgassing of oxidised species such as $\mathrm{CO_2, SO_2}$ and endogenously produced $\mathrm{H_2O}$. In the second pathway, vigorous magma ocean turbulence inhibits efficient iron segregation, preserving more reducing conditions inherited from accretion and favouring the outgassing of reduced species such as $\mathrm{H_2}$ and $\mathrm{CH_4}$. The coupling between mantle geochemistry and atmospheric composition suggests that the secondary atmospheres of rocky planets can broadly be separated into reduced, oxidised, and intermediate classes \citep{Liggins_2022}. The oxygen fugacity ($f\mathrm{O_2}$), an effective partial pressure of O$_2$, serves as a proxy for the mantle redox state because it quantifies the relative abundance of valence states in polyvalent elements such as iron \citep{Frost_1991, Gaillard_2014, Ortenzi_2020, Sossi_2020}. It therefore controls the speciation and partitioning of volatiles within the magma ocean \citep{Schaefer_2017}.

Photochemistry refers to chemical reactions initiated or altered by radiation, and, in the context of exoplanet atmospheres, is a disequilibrium process that couples stellar irradiation to atmospheric chemistry. By altering molecular abundances and producing reactive intermediates, photochemistry interacts with outgassing, radiative transfer, and vertical transport and can therefore influence both the structure of secondary atmospheres and the interpretation of their observable spectra \citep{Marley_2008, Seager_2010, Linsky_2014, Moses_2014}. Chemical kinetics models provide a framework for exploring the effects of such processes in a self-consistent way \citep{Venot_2012, Tsai_2021}. Most coupled rocky-planet evolution frameworks simplify atmospheric chemistry, particularly photochemical processes, and rely on equilibrium chemistry, reduced volatile networks, or simplified photochemical treatments to model long-term planetary evolution \citep{Ortenzi_2020, Liggins_2022, Krissansen_Totton_2022, Nicholls_2024b, Maurice_2024}. Stellar irradiation affects the composition of the atmosphere at the level probed by transmission spectroscopy with instruments such as the \textit{James Webb} Space Telescope (JWST) \citep{Madhusudhan_2019}. It is therefore essential that its effects are extensively explored. 

An example of the significance of photochemical reactions can be found in the atmosphere of the hot Jupiter WASP--39b. The detection of $\mathrm{SO_2}$ via a spectral signature at $4\,\mu$m from JWST could be explained by oxidising $\mathrm{H_2S}$ to $\mathrm{SO_2}$ via photolysis \citep{Tsai_2023}. Additionally, \citet{Dai_2026} recently presented a real example of complex photochemistry acting on a system with two young Jupiter-sized planets, V1298~Tau\,e and b, producing two different signatures through the detection of $\mathrm{CS_2}$ for the former and $\mathrm{SO_2}$ for the latter. Various chemical kinetics models that include high-temperature reactions \citep{Moses_2011}, extensive networks validated with combustion mechanisms \citep{Venot_2012}, effects from sulfur species \citep{Hobbs_2021}, and the formation of aerosols through photochemical haze precursors \citep{Steinrueck_2021} have been developed to isolate photochemical fingerprints on hot Jupiters and their observational spectra \citep{nicholls_temperature_2023}. 

Photochemical effects have been partially explored in sub-Neptunes and super-Earths, where the question of whether disequilibrium species can provide insight into planetary interiors and outgassing processes has been proposed. Previous studies have shown that photochemically fragile species are sensitive to interior volatile composition, atmospheric mixing, and stellar irradiation, allowing atmospheric observations to place constraints on interior chemistry and planetary evolution \citep{Yu_2021, Hu_2021, Tsai_2021, Hakim_2026}. The case of L~98--59\,d indicates $\mathrm{SO_2}$ in the upper atmosphere \citep{Gressier_2024, Banerjee_2024}, which cannot be explained by volcanic outgassing. The low-density magma ocean super-Earth \citep{Cadieux_2025} is rich in sulfur, but a volcanic explanation for the high-altitude presence of $\mathrm{SO_2}$ would require a rise from the surface through a rich $\mathrm{H_2}$ background without thermochemical reduction. \citet{Nicholls_2026} propose that photochemical kinetics can explain its presence despite the large uncertainties in the JWST observations of this planet. Similarly, \citet{Tsai_2026} show that coupled climate and photochemical modelling can reproduce the observed spectrum of sub-Neptune K2--18\,b without requiring dimethylsulfide. Their work illustrates that photochemical products are key tracers in distinguishing different habitable conditions and interior structures of K2--18\,b. 

Among volatile species of interest, sulfur-bearing molecules are particularly promising because they are sensitive to both the redox state of the mantle and atmospheric photochemistry, making them suitable tracers of interior evolution and volcanic outgassing on rocky exoplanets \citep{Zahnle_2009, Kaltenegger_2010, Hu_2012, Gaillard_2014, Tsai_2023}. Sulfur enrichment of rocky planets may arise from the protoplanetary disk through ammonium salts such as $\mathrm{NH_4SH}$ that are accreted during their formation and processed by photochemistry \citep{Nakazawa_2026}. The abundance of sulfur gases is theorised to be inversely coupled to the host star's UV flux and can help trace the habitable zone of M-dwarf stars \citep{Jordan_2025}. Sulfur chemistry spans photochemical processing, oxidation, and solubility, with effects that can be directly observed in the planetary context due to its strong absorption features. Sulfur can adopt multiple oxidation states, causing its speciation in planetary interiors and outgassed atmospheres to vary strongly with oxygen fugacity and chemical environment \citep{Jugo_2010, Gaillard_2014}. On Earth, volcanically outgassed $\mathrm{SO_2}$ highlights physical and chemical processes in the interior \citep{Carn_2017}, with sulfur oxidation to $\mathrm{SO_2}$ occurring at water-rich subduction zones \citep{Wallace_2001}.

Aerosol precursors in the sulfur network can drive sulfate aerosol formation in the atmosphere \citep{Gao_2017}. Aerosols are microscopic particles suspended in the atmosphere that can affect the climate by scattering and absorbing incoming stellar radiation, thereby modifying the planet’s radiative balance \citep{Deming_2017, Gao_2021}. In exoplanet observations, sulfur-bearing aerosols are particularly important because they can modify transmission spectra, mute or obscure molecular absorption features, and alter a planet’s reflected and emitted flux, thereby complicating the interpretation of transit and eclipse measurements \citep{Gao_2017, Gao_2021}.

In this work, we investigate how photochemistry modifies the coupling between interior redox evolution and atmospheric composition across different classes of rocky planets during the magma ocean phase. By combining coupled interior-atmosphere evolution models with chemical kinetics calculations, we explore whether photochemical disequilibrium processes significantly alter atmospheric compositions predicted from equilibrium outgassing alone, particularly for sulfur-bearing species that may act as tracers of mantle oxidation state. We further examine how these effects influence the observable properties of rocky exoplanet atmospheres and their interpretation with current and future spectroscopic observations. In Sect. \ref{sec: Simulation Setup}, we describe the coupled atmosphere-interior framework and the atmospheric chemistry modelling. In Sect. \ref{sec: Results}, we present the bulk atmospheric compositions, the effects of photochemistry on the upper parts of the atmosphere, the behaviour of sulfur species, and notable observational imprints. Section \ref{sec: Discussion} discusses implications, comparison with previous studies, simplifying assumptions made in our work, and suggestions that could further affect atmospheric characterisation. Finally, we summarise our conclusions in Sect. \ref{sec: Conclusions}.

\section{Simulation setup}\label{sec: Simulation Setup}
\subsection{Coupled interior-atmosphere evolution framework}
The \texttt{PROTEUS} framework \citep{Lichtenberg_2021b, Nicholls_2024a} is used to simulate the evolution of rocky planets after the accretion of their primary atmosphere. It is a coupled modular simulation framework that primarily combines the 1D interior dynamics module for rocky planets made of molten and solid interiors, \texttt{SPIDER} \citep{Bower_2018, Bower_2019}, with the atmospheric radiative-convective solver, \texttt{AGNI} \citep{Nicholls_2024b, Nicholls_2025}. In this work, an early Earth-sized planet was simulated. An initial planetary inventory of C-H-O-N-S volatiles was set to $\mathrm{C/H} = 1.0, \mathrm{N/H} = 0.5$, and $ \mathrm{S/H} = 2.0$, with a total amount of hydrogen equivalent to five Earth oceans \citep{Korenaga_2017}. These values are intended to represent an Earth-like volatile inventory for a water-rich rocky planet \citep{Peslier_2017, Wang_2018} and are broadly consistent with the initial conditions adopted in previous magma-ocean and coupled interior-atmosphere evolution studies of terrestrial planets \citep{Lichtenberg_2021, Liggins_2022, Nicholls_2024a}.

During the simulated evolution of the planet, volatiles are distributed among the partially molten mantle and the surrounding atmosphere through equilibrium outgassing and thermochemical reactions. Importantly, the abundance of oxygen is controlled by the iron-w\"ustite buffer ($\mathrm{IW}$) reaction ($\mathrm{2Fe + O_2 \rightleftharpoons 2FeO}$), as a function of temperature \citep{Oneill_2002}. \texttt{SPIDER} evolves an Earth-like magma ocean, adopting the equation of state of a pure $\mathrm{MgSiO_3}$ composition. The simulated mantle is initialised fully molten on an adiabatic temperature profile and then evolves according to radial energy fluxes between each layer, which describe the transfer of heat between concentric layers as a function of planetary radius. \texttt{SPIDER} accounts for energy transport with mixing-length convection, in addition to latent heat of crystallisation, phase separation, and grain gravitational settling \citep{Bower_2018, Bower_2019}. A fixed interior structure is assumed, with disabled radiogenic and tidal heating effects.

Our atmosphere climate model, \texttt{AGNI}, calculates atmospheric energy fluxes based on the incident stellar radiation, the volatile composition outgassed from the magma ocean, and the (time-evolving) temperature at the top layer of the mantle. The \texttt{SOCRATES} radiative transfer code is used to solve for radiation fluxes at each atmospheric layer, based on correlated-$k$ two-stream radiative transfer with 48 spectral bands and Rayleigh scattering \citep{Edwards_1996, Amundsen_2014}. The atmosphere's convective energy flux is parametrised using canonical mixing-length theory \citep{Joyce_2023}. Simultaneously with planetary evolution, stellar evolution is also modelled. The stellar radius, bolometric luminosity, X-ray emission, and UV emission are calculated using a two-shell rotational evolution model based on empirically derived scaling laws \citep{Johnstone_2021}. Atmospheric escape is included in the framework but was not enabled here. The magma-ocean evolution of the Earth-like planets modelled with \texttt{PROTEUS} is terminated based on the following convergence criteria: global energy balance or mantle crystallisation. The former quasi-steady state means that the total outgoing radiative flux of the simulated planet, combined with the heat supplied from the interior to the surface and atmosphere, matches the absorbed stellar flux within numerical tolerance. The latter is reached when the mantle melt fraction, or the ratio between the liquid and the total material in the mantle, drops below 0.01.

\subsection{Atmospheric chemistry and photochemical modelling}\label{sec: atmospheric chemistry}
The three regimes explored in this work are equilibrium, non-equilibrium without photochemistry (only vertical transport), and non-equilibrium with photochemistry. To model disequilibrium atmospheric composition, the final atmospheric structures produced by \texttt{PROTEUS} are post-processed using \texttt{FastChem} and \texttt{VULCAN}. The chemistry modules \texttt{FastChem} and \texttt{VULCAN} simulate the effects of equilibrium and disequilibrium chemistry, respectively. The outgassing abundances, along with the pressure-temperature profile of the final time step, act as boundary conditions for the equilibrium state of the atmosphere. The equilibrium state refers to thermochemical equilibrium, which occurs when there is no further driving force for heat flow, mass transfer, or chemical reactions for all the species participating in reactions between two mixtures at the same temperature and pressure \citep{DeVoe_2020}. \texttt{FastChem} solves the system of mass action law equations for the gas phase by using elemental abundances and minimising the Gibbs free energy \citep{Heng_2016b, Woitke_2018}. The law of mass action states that, at chemical equilibrium, the abundances of reactants and products are related through equilibrium constants, while elemental conservation fixes the total inventory of each element \citep{Denbigh_1955, Stock_2018}. Elemental abundances are extracted by stoichiometrically breaking down all the included species in the atmosphere and adding them together, as input \citep{Stock_2018, Kitzmann_2023}. Thermochemical reactions drive the gas towards equilibrium, so the composition is primarily controlled by local temperature, pressure, and elemental abundances \citep{Wordsworth_2022}.

The equilibrium state calculated with \texttt{FastChem} acts as the initial state for \texttt{VULCAN}'s kinetics network, which was originally described in \citet{Tsai_2017}. In this work, the C-H-O-N-S photochemical network is implemented. \texttt{VULCAN} includes advection, eddy diffusion, and molecular diffusion to describe atmospheric transport. The three components of the transport flux equation are the following:
\begin{equation}
\phi_{\rm i} =
- n_{\rm i} v
- K_{\rm zz} n_{\mathrm{tot}} \frac{\partial X_{\rm i}}{\partial z}
- D_{\rm i} \left[\frac{\partial n_{\rm i}}{\partial z} + n_{\rm i}\left(
\frac{1}{H_{\rm i}},
+,\frac{1+\alpha_{\rm T}}{T}, \frac{dT}{dz}\right)
\right],
\label{eq:transport_flux}
\end{equation}
where $v$ is the vertical wind velocity, $K_{\rm zz}$ and $D_{\rm i}$ are the eddy and molecular diffusion coefficients, respectively. Moreover, $H_{\rm i}$ is the molecular scale height for species i with molecular mass $\mu_{\rm i}$, expressed as $H_{\rm i} = k_{\rm B}T/\mu_{\rm i} g$, where $g$ is the gravitational acceleration, $T$ is the temperature, and $k_{\rm B}$ is the Boltzmann constant. The parameter $\alpha_{\rm T}$ is the thermal diffusion factor \citep{Tsai_2021}. Eddy diffusion represents the 3D process of vertical transport by turbulent or convective motions not resolved in 1D, tending to homogenise the atmospheric composition over a characteristic length \citep{Gao_2021}. Molecular diffusion describes individual species and their microscopic transport caused by concentration gradients, gravitational separation, and thermal diffusion. It usually becomes important when the atmosphere is thin enough that species can separate by molecular mass \citep{Moses_2014}. Equation \ref{eq:transport_flux} assumes hydrostatic balance and uses the eddy diffusion coefficient $K_{\rm zz}$ directly from \texttt{AGNI}'s radiative-convective solution to self-consistently model the aforementioned vertical transport effects, referred to as mixing. \texttt{AGNI} uses mixing length theory to calculate $K_{\rm zz} = \lambda_{\rm conv} w_{\rm conv}$, where $\lambda_{\rm conv}$ is the mixing length and $w_{\rm conv}$ is the convective velocity \citep{Nicholls_2025}. 

The inclusion of photochemical reaction pathways to \texttt{VULCAN} is explained in \citet{Tsai_2021}. The final model input is the stellar spectrum, which matches the evolutionary step of the planet. \texttt{VULCAN} simulates the short-wave actinic radiation flux, which determines the rate of photoreactions throughout the atmosphere. The efficiency of this process depends on the overlap between the stellar radiation field, given by the actinic flux $F(\lambda)$, the molecular absorption cross section $\sigma(\lambda)$, and the photodissociation quantum yield $\phi(\lambda)$, which gives the fraction of absorbed photons that actually lead to dissociation \citep{Noelle_2020}. The photolysis rate ($J$) quantifies the speed of photodissociation. It is represented as an integral over the wavelength range as \citep{Schinke_1993, Balis_2002}:
\begin{equation}
    J = \int \sigma(\lambda)\,\phi(\lambda)F(\lambda)\,d\lambda.
\end{equation}
The actinic flux includes attenuation of the downwelling stellar beam with diffusely scattered and absorbed components. The relevant absorption cross sections are extracted from the Leiden Observatory database \citep{Heays_2017}.

\subsection{Simulation grid and atmospheric characterisation}
We explored the effects of photochemistry across a range of rocky-planet environments by varying the mantle redox state, stellar irradiation, and stellar spectral type. Fifty simulations were performed with \texttt{PROTEUS} using a grid of three parameters. Oxygen fugacity (gridded as log $f\mathrm{O_2}$, relative to IW) quantifies the mantle redox state and controls volatile speciation and partitioning in the magma ocean. Instellation flux is the total amount of irradiation a planet receives from its host star. The present-day solar constant for Earth is approximately $S_\odot =1360\,\mathrm{W\,m^{-2}}$ \citep{Prsa_2016}. The oxygen fugacity and the instellation flux are varied as $\log (f\mathrm{O_2})=$ IW-4, IW-2, IW+0, IW+2, IW+4, and  $0.1\,S_\odot, 1.0\,S_\odot, 10\,S_\odot, 100\,S_\odot$, and $1000\,S_\odot$, respectively. The third parameter is the stellar spectrum, with the Sun's and GJ 1132's being used. The latter is referred to simply as `M-dwarf' in the following sections. GJ 1132 is a small M-dwarf type star with a mass of $0.194\,M_\odot$ and a bolometric luminosity of $0.00436\,L_\odot$ \citep{Pineda_2021}. Its spectrum is used as a representative active M-dwarf irradiation environment, which enables the isolation of enhanced X-ray and UV emission effects on atmospheric photochemistry, compared to the Sun's irradiation spectrum at the same bolometric flux. The stellar spectra were adopted from the spectral database distributed with \texttt{MORS}, which primarily derives stellar UV-IR spectra from the MUSCLES survey and associated reconstructed stellar spectra \citep{France_2016, Youngblood_2016}. The solar spectrum follows \citet{Gueymard_2004}, while the GJ 1132 spectrum is based on reconstructed UV observations \citep{Waalkes_2019}. For the Sun, the instellation flux values in increasing order correspond to the approximate distances of 0.03, 0.1, 0.3, 1.0, and $3\,\mathrm{au}$, whereas for GJ 1132, they correspond to 0.002, 0.007, 0.02, 0.07, and $0.2\,\mathrm{au}$. All the parameters explored in our study are shown in Table \ref{tab:proteus_grid}.

\begin{table}[t]
\centering
\caption{Overview of the parameter space explored in this study.}
\begin{tabular}{lc}
\hline \hline
Parameter & Values \\ \hline
$\log(f\mathrm{O_2})$, relative to IW & $-4$, $-2$, $+0$, $+2$, $+4$ \\ 
Instellation flux, $S_\odot$ & $0.1$, $1.0$, $10$, $100$, $1000$ \\ 
Stellar spectrum & Sun, GJ 1132 (M-dwarf) \\ \hline
\end{tabular}
\label{tab:proteus_grid}
\end{table}

For each simulation, three post-processing atmospheric characterisations were constructed: the equilibrium result from \texttt{FastChem}, a disequilibrium case with only mixing effects, and a disequilibrium case with photochemistry and mixing combined. This approach allows for the separation of photochemical effects from vertical transport, namely the two disequilibrium processes explored in this work. The final outputs include the volume mixing ratios (VMRs) of 85 species, which is the total number normalised over the entire 41-layer atmosphere. Finally, to assess how the disequilibrium chemistry affects observables, selected atmospheric profiles were post-processed with the radiative transfer module \texttt{AGNI} to generate medium-resolution synthetic spectra, using 4069 spectral bands \citep{Nicholls_2024b}. For each case, the pressure-temperature structure and vertical volume mixing ratio profiles of all chemically active species were exported from the atmospheric calculation and supplied to \texttt{AGNI} as a fixed background atmosphere. \texttt{AGNI} then computed wavelength-dependent short-wave and long-wave radiative fluxes using a medium-resolution spectral configuration that includes Rayleigh scattering and gaseous continuum opacity \citep{Nicholls_2025}. For the wavelength range 0.6 to $15\,\mu$m, the spectral resolution (R value) is between 725 and 740, covering 2355 wavelength bins. From these calculations, the top-of-atmosphere spectral fluxes and contribution functions were extracted, which together provide a direct measure of how photochemistry and vertical mixing modify the spectrally active layers of the atmosphere.

\section{Results}\label{sec: Results}

\subsection{Bulk atmospheric compositions}\label{subsec: overall}

\begin{figure}[h!]
    \centering
    \includegraphics[width=\columnwidth]{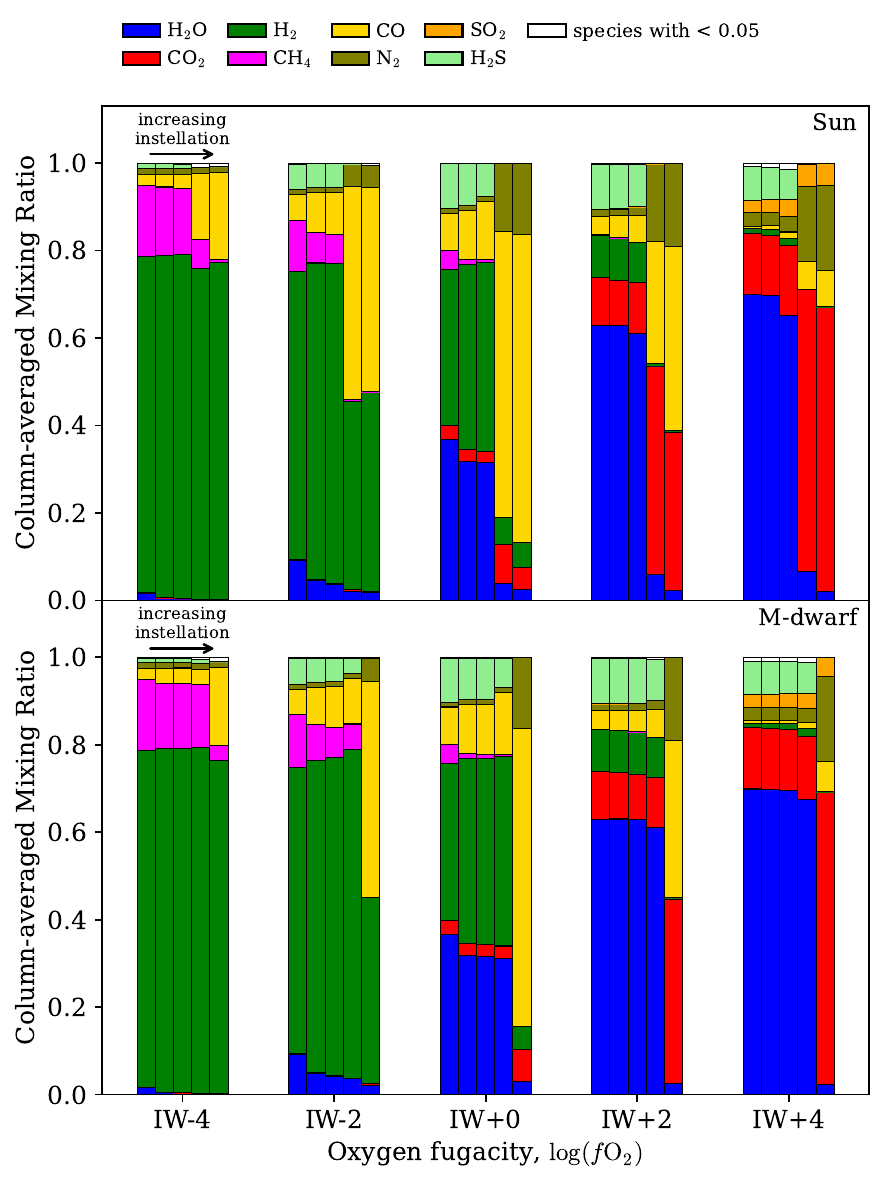}
    \caption{Normalised column-integrated abundances for the dominant atmospheric species over all mantle redox states and instellation fluxes for the Sun (top) and M-dwarf (bottom) photochemistry-plus-mixing cases. Each mantle redox state (the IW value) includes five vertical bars corresponding to the instellation flux values of 0.1, 1.0, 10, 100, and $1000\,S_\oplus$, increasing from left to right. The dominant volatiles are indicated in different colours, and the species with a column-averaged mixing ratio less than 0.05 are shown in white.}
    \label{fig:cdrs}
\end{figure}
\begin{figure}[h!]
    \centering
    \includegraphics[width=\columnwidth]{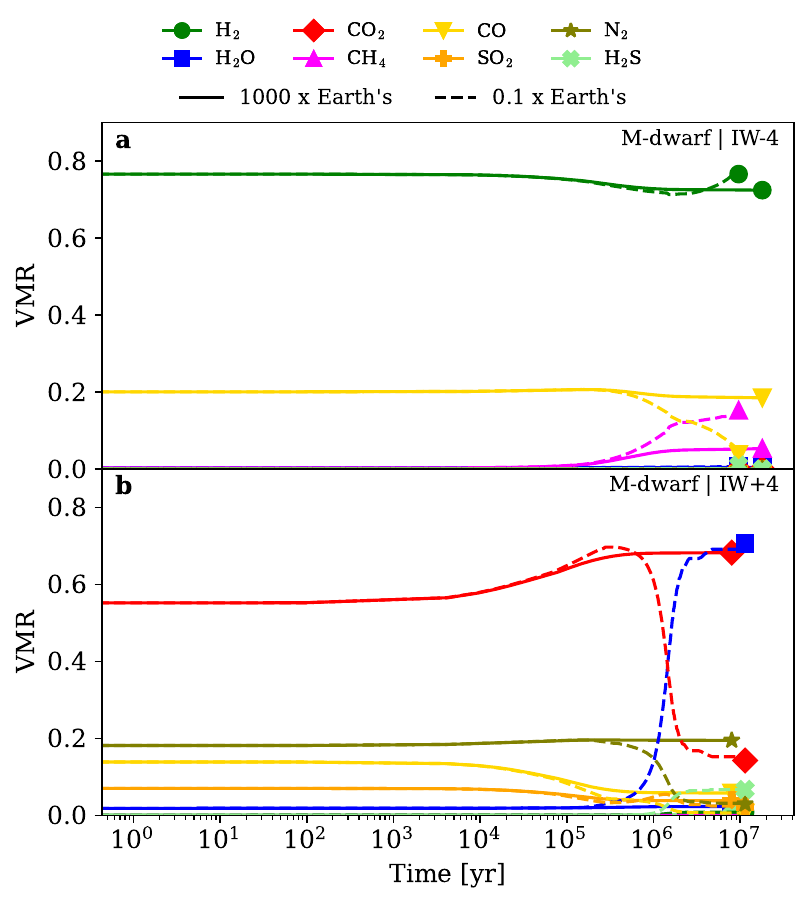}
    \caption{Evolution of the outgassed major volatile abundances for the most reduced (top) and most oxidised (bottom) M-dwarf cases at the lowest and highest instellation fluxes. The markers indicate the final outgassed abundances used as inputs for the post-processed atmospheric characterisations. The volume mixing ratio values are normalised over all outgassed species.}
    \label{fig:outgassing_01vs1000}
\end{figure}
The bulk atmospheric compositions across all three atmospheric characterisations were compared using species column-integrated abundances, which provide a measure of the total abundance of each outgassed species. For each simulation, the column-averaged mixing ratio abundance was calculated by multiplying the volume mixing ratio of each species by pressure, dividing by temperature, and summing the result across all atmospheric layers. This approach enables a direct comparison of atmospheric composition across different mantle redox states, instellation fluxes, and (dis)equilibrium chemistry scenarios. Figure \ref{fig:cdrs} shows a strong dependence of atmospheric composition on mantle redox state, with highly reduced cases, $\mathrm{IW}$-4, dominated by $\mathrm{H_2}$ and $\mathrm{CH_4}$. More oxidised cases show enhanced $\mathrm{H_2O}$ and $\mathrm{CO_2}$ abundances, while intermediate redox states exhibit a peak in $\mathrm{CO}$. Species such as $\mathrm{H_2S}$ and $\mathrm{N_2}$ remain roughly independent of the redox state and are present over all cases. The overall bulk atmospheric composition remains broadly unchanged between the equilibrium, mixing-only, and photochemistry-plus-mixing atmospheric characterisations, with no significant differences visible in any of the three versions of Fig. \ref{fig:cdrs}, which only shows the result for the latter disequilibrium case. Photochemical processing primarily redistributes species in the upper atmosphere without substantially altering the total volatile inventory.

\begin{figure}[h!]
    \centering
    \includegraphics[width=\columnwidth]{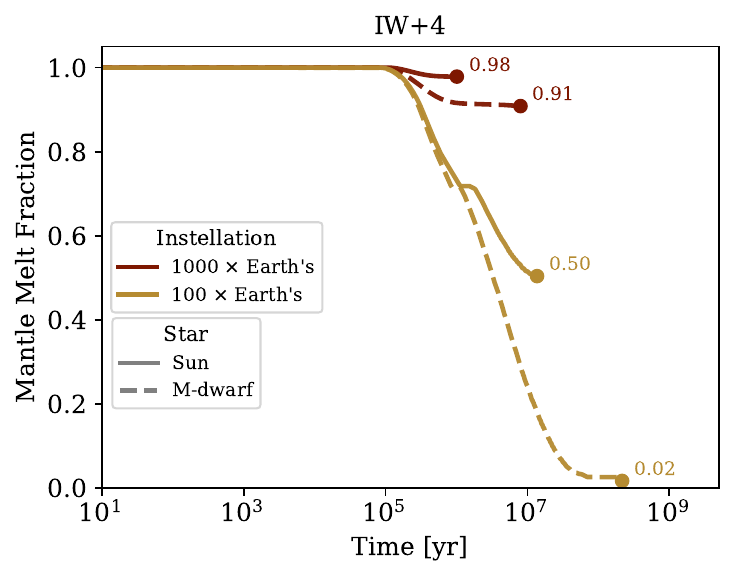}
    \caption{Evolution of the mantle melt fraction for the two highest instellation fluxes (100 and $1000\,S_\oplus$) with $\mathrm{IW}$+4, for the Sun (solid) and the M-dwarf (dashed) cases. The value at the final time step is indicated next to the marker.}
    \label{fig:mantle_solidification}
\end{figure}
\begin{figure}[h!]
    \centering
    \includegraphics[width=\columnwidth]{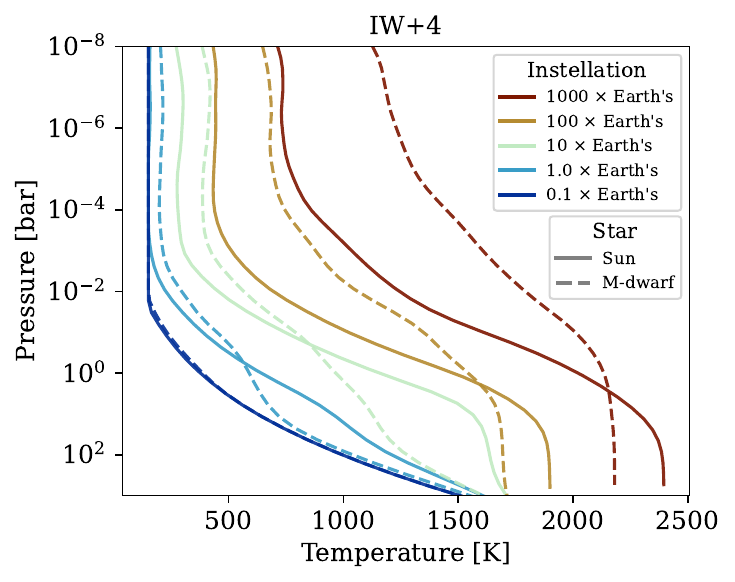}
    \caption{Pressure-temperature profiles at the final converged evolutionary time step for the $\mathrm{IW}$+4 cases. The colour indicates the instellation flux, where the solid lines correspond to the Sun cases and the dashed lines to the M-dwarf cases.}
    \label{fig:pt_profiles}
\end{figure}

For a given redox state, differences in the outgassed composition are associated with the outgassing history and final convergence state. For cases with lower instellation fluxes (first three vertical bars across all mantle redox states for both stellar cases in Fig. \ref{fig:cdrs}), the mantle solidified, resulting in efficient outgassing of volatiles such as $\mathrm{H_2O}$. Figure \ref{fig:outgassing_01vs1000} shows the time evolution of the outgassing volatile abundances for the most reduced, IW-4, and most oxidised, IW+4, cases, across both their lowest and highest instellation fluxes, 0.1 and $1000\,S_\oplus$. Figure \ref{fig:outgassing_01vs1000} \textbf{a} and \textbf{b} correspond to the first and fifth bars of IW-4 and IW+4 in Fig.\ref{fig:cdrs}, respectively. Water is much more soluble in silicate melt compared to a solid mineral lattice, allowing it to enrich the melt until saturation \citep{Ottonello_2018}. As a result, panel \textbf{b} of Fig. \ref{fig:outgassing_01vs1000} shows a spike in the volume mixing ratio of outgassed $\mathrm{H_2O}$ from essentially 0.0 to around 0.7, as the planet evolved between $2\times 10^5$ and $10^7$ yr. The volume mixing ratio of $\mathrm{CO_2}$ and other indicated volatiles decreases due to normalisation over all species. A similar trend is observed for $\mathrm{CH_4}$ (shown in panel \textbf{a} of Fig \ref{fig:outgassing_01vs1000}). For reducing cases, the increase in $\mathrm{CH_4}$ is attributed to thermochemical equilibrium reactions between C and H species in the mantle, rather than solubility. 

The outgassed compositions of the M-dwarf cases shown in Fig. \ref{fig:outgassing_01vs1000} evolve similarly to the equivalent Sun cases. More irradiated cases (100 and $1000\,S_\oplus$) reach global energy balance with permanent magma oceans, while less irradiated cases (0.1, 1.0, and $10\,S_\oplus$) have a solidified mantle. However, for the cases with $100\,S_\oplus$ (every mantle redox state's fourth bar in the bottom panel of Fig. \ref{fig:cdrs}), the different outgassing history changes the bulk atmospheric composition across all mantle redox states. In these simulations, the planet still converges to a global energy balance, but only after a substantial fraction of the mantle has solidified, releasing a significant amount of $\mathrm{H_2O}$ through outgassing. The same trend is seen for all redox states at $100\,S_\oplus$  for the M-dwarf cases, with mantle melt fractions ranging from 0.02 to 0.04. By contrast, the corresponding Sun cases at the same instellation retain mantle melt fractions of about 0.50, which is too large for mantle solidification to significantly modify the bulk atmospheric composition. The mantle melt fraction between the Sun and M-dwarf IW+4 cases for 100 and $1000\,S_\oplus$ is shown in Fig. \ref{fig:mantle_solidification} as a representative example.

Another difference between the results of the two stellar hosts comes from their influence on the magma ocean's atmospheric pressure-temperature profile. For the same bolometric flux, the M-dwarf can heat the upper atmosphere significantly more compared to the Sun. Figure \ref{fig:pt_profiles} shows the pressure-temperature profiles for the last IW+4 snapshots over all instellation fluxes for both the Sun and the M-dwarf to illustrate their differences, as calculated by the radiative-convective solver \texttt{AGNI}. The surface pressure is calculated around $10^3$ bar over all cases, as predicted for the early magma ocean Earth \citep{Zahnle_2010}. As instellation increases, the differences in the temperature of the upper atmosphere increase, with the atmospheres subject to the M-dwarf's irradiation reaching 1250\,K at $10^{-8}$ bar compared to 800\,K for the Sun at $1000\,S_\oplus$. The surface temperature follows an opposite trend, with 2200\,K for the M-dwarf cases compared to 2400\,K for the Sun. The M-dwarf radiation is distributed towards longer wavelengths, so the $\tau=1$ absorption level occurs at lower pressures, higher in the atmosphere. Solar radiation penetrates deeper, driving convection, inducing a steeper lapse rate, and resulting in higher surface temperatures \citep{Malik_2019}. Overall, the bulk atmospheric compositions are primarily controlled by mantle redox state and volatile outgassing history, while stellar spectral type and photochemistry mainly modify the upper atmospheric thermal structure, as explained in Sect. \ref{subsec: top of atm}.

\subsection{Photochemical effects at the top of the atmosphere}\label{subsec: top of atm}

\begin{figure*}[h!]
    \centering
    \includegraphics[width=\textwidth]{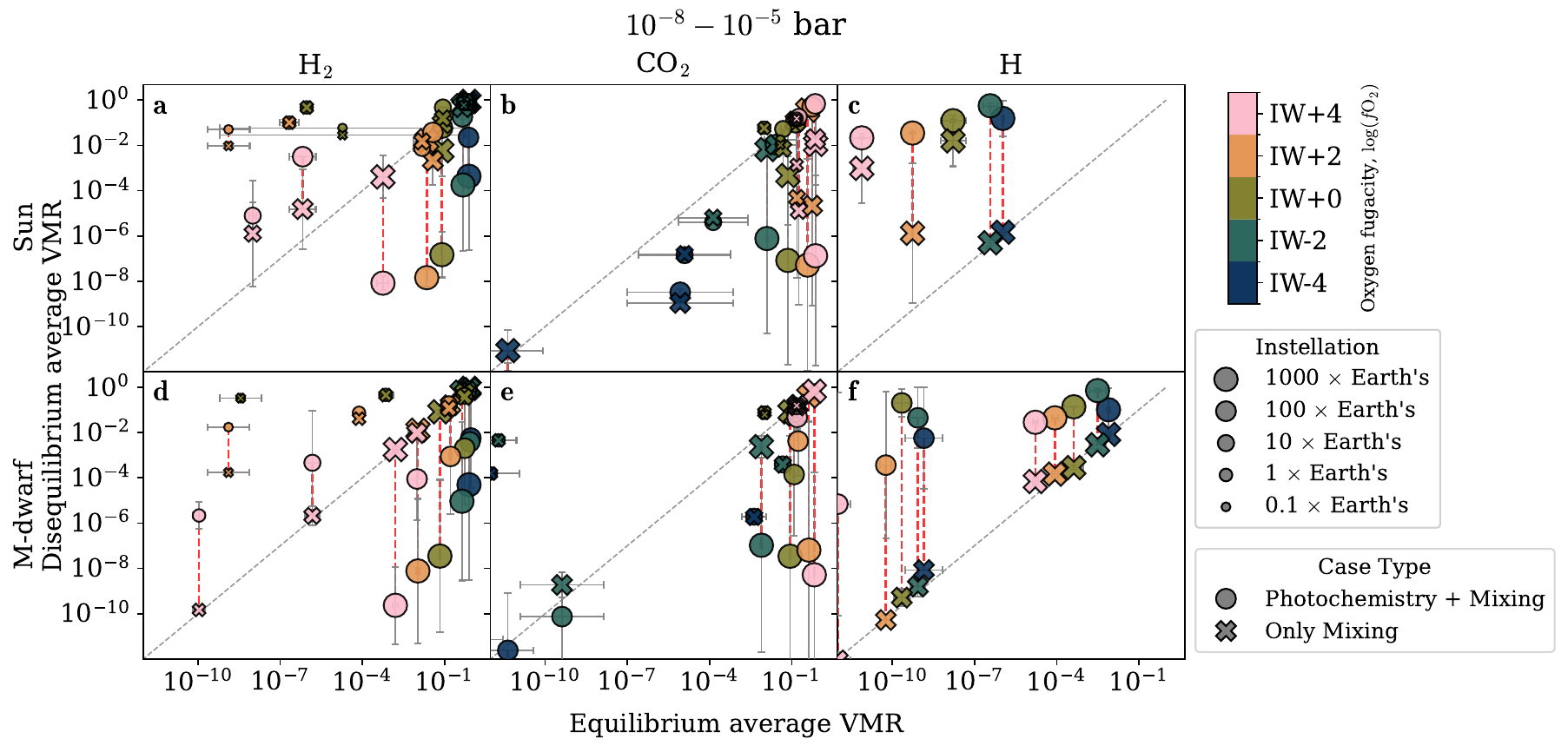}
    \caption{Comparison of the species abundances near the top of the atmosphere after the equilibrium and disequilibrium analysis for the Sun and M-dwarf cases. Each panel shows one species ($\mathrm{H_2, CO_2}$, and H), with the x-axis giving the equilibrium log-mean volume mixing ratio and the y-axis giving the corresponding disequilibrium value averaged over $10^{-8}$--$10^{-5}$ bar. The error bars represent the standard deviation divided by the number of points for the volume mixing ratios between these two pressure layers. The circles denote the result for mixing plus photochemistry, while the X symbols denote the result with only mixing. The dashed red connectors link points from the same case at a fixed equilibrium abundance to distinguish the photochemical from the mixing results. The marker colour encodes the mantle redox state $\log (f\mathrm{ O_2})$, the marker size encodes instellation ($0.1$--$1000\,S_\oplus$), and the dashed black diagonal marks the 1:1 agreement with equilibrium. Deviations from the diagonal quantify departures from thermochemical equilibrium, with the upper-left part corresponding to enhancement and the lower-right part corresponding to depletion.}
    \label{fig:upper_effects}
\end{figure*}

We now consider the effects of photochemistry and mixing in driving atmospheric compositions out of equilibrium at the upper part of the atmosphere. Figure \ref{fig:upper_effects} plots final species abundances post-processed with \texttt{VULCAN} (only mixing, photochemistry plus mixing) and \texttt{FastChem} (thermochemistry). Photochemistry strongly modulates the uppermost atmosphere compositions, across all simulation cases with $1000\,S_\oplus$, through significant photodissociation of molecular volatiles. Figure \ref{fig:upper_effects} highlights this effect for $\mathrm{H_2}$ (\textbf{a, d}) and $\mathrm{CO_2}$ (\textbf{b, e}), which are the dominant species in the reduced and oxidised cases, respectively. The logarithmic average of the volume mixing ratio between $10^{-5}$ and $10^{-8}$ bar deviates from equilibrium, which is indicated with the diagonal dashed line, most substantially for the highest instellation fluxes, 100 and $1000\,S_\oplus$. Stellar irradiation causes the production of a notable amount of radicals, with H (shown in panels \textbf{c, f}) being the most substantial in number and showing the most significant changes in abundance.

For highly reduced cases with IW-4 and IW-2, $\mathrm{H_2}$ dominates the bulk atmospheric composition, but stellar irradiation above $10\,S_\oplus$ induces photodissociation, which increases in strength with redox state and instellation flux. For the Sun cases with $1000\, S_\oplus$ flux, a rough trend can be observed in Fig. \ref{fig:upper_effects} \textbf{a} that decreases the volume mixing ratio of $\mathrm{H_2}$ from approximately $10^{0}$ to $10^{-3.5}$ for IW-4 and from $10^{-3}$ to $10^{-8}$ for IW+4. The same behaviour is observed for the M-dwarf cases, but the stronger X-ray and extreme ultraviolet (XUV) radiation enhances the effect from approximately $10^{0}$ to $10^{-4.5}$ for IW-4 and from $10^{-3}$ to $10^{-10}$ for IW+4, as shown in Fig. \ref{fig:upper_effects} \textbf{d}. A similar trend is also seen for $100\,S_\oplus$, which does not appear for the Sun (decrease from $10^0$ to $10^{-2}$ for IW-4, and from $10^{-2}$ to $10^{-4}$ for IW+4 in Fig. \ref{fig:upper_effects} \textbf{d}). For a few more oxidised cases with lower instellation flux, stellar irradiation declines to a degree that enables a slight enhancement of $\mathrm{H_2}$ due to recombination. This can be seen in the Sun case with IW+4 where its average volume mixing ratio increases from $10^{-5}$ to $10^{-2.5}$ for 100 $S_\oplus$ (Fig. \ref{fig:upper_effects} \textbf{a}), and for the M-dwarf case with IW+4 where it increases from $10^{-6}$ to $10^{-3}$ for 10 $S_\oplus$ (Fig. \ref{fig:upper_effects} \textbf{d}). For even lower irradiation, enhancement in the upper atmosphere can be attributed to mixing, which transports $\mathrm{H_2}$ from the deeper layers and can increase the volume mixing ratio by approximately 3 to 8 orders of magnitude (see cases with $0.1-1.0\,S_\oplus$ and IW+0, IW+2 in Fig. \ref{fig:upper_effects} \textbf{a, d}).

For highly oxidised cases with IW+2, IW+4, the bulk composition is dominated by $\mathrm{CO_2}$ except for the lower instellation flux cases where mantle solidification allows a significant exsolution of $\mathrm{H_2O}$, as described in Sect. \ref{subsec: overall}. $\mathrm{CO_2}$ exhibits a similar behaviour to $\mathrm{H_2}$ for $1000\,S_\oplus$ with very strong photodissociation, and follows the same pattern of more significant photodissociation with increasing mantle redox state. Figure \ref{fig:upper_effects} \textbf{b} shows a decrease from $10^{-2}$ to $10^{-6}$ for IW-4, and a decrease from $10^{-2}$ to $10^{-7}$ for IW+4. Similarly, Fig. \ref{fig:upper_effects} \textbf{e} shows a decrease from $10^{-3}$ to $10^{-7}$ for IW-4, and a decrease from $10^{-1}$ to $10^{-8}$ for IW+4. In contrast to $\mathrm{H_2}$, mixing mostly transports more $\mathrm{CO_2}$ from the upper part to the deeper part of the atmosphere, as can be seen by the large number of low-instellation "$\times$" markers in the lower-right part of Fig. \ref{fig:upper_effects} \textbf{b} (Sun cases). The effect is less obvious for the M-dwarf cases in Fig. \ref{fig:upper_effects} \textbf{e}.

The highly irradiated part of the atmosphere is populated with H radicals for both the Sun and M-dwarf cases. Other radicals, such as O and C, are also present, but with lower volume mixing ratios compared to H (less than $10^{-12}$). The trend observed in $\mathrm{H_2}$ can also be seen for H with an increase from about $10^{-6}$ to $10^{-1}$ for IW-4 and from $10^{-3}$ to $10^{-2}$ for IW+4 in Fig. \ref{fig:upper_effects} \textbf{c} ($1000\,S_\oplus$ for the Sun). For the M-dwarf cases in Fig. \ref{fig:upper_effects} \textbf{f}, the trend also appears for 100 $S_\oplus$, reflecting the behaviour of $\mathrm{H_2}$ in Fig. \ref{fig:upper_effects} \textbf{d} and the effects of stronger UV irradiation. The maximum enhancement of H abundance can be seen for IW+0, from about $10^{-9}$ to $10^{-1}$.

Figure \ref{fig:outgassing_01vs1000} shows that the outgassing timescale of our rocky planets lies in the range of $10^6$ and $10^7$ yr. To identify the extent of its effect and how it is related to the dominant processes controlling the upper atmosphere, we compute and compare its values with two characteristic timescales at each pressure level: mixing and photolysis. The mixing timescale is defined as $\tau_{\rm mix} = H^2/K_{\rm zz}$, where $H$ is the pressure scale height and $K_{\rm zz}$ is the eddy diffusion coefficient \citep{Zhang_2018}. For the most highly irradiated cases ($1000\,S_\oplus$), $\tau_{\rm mix}$ typically lies in the range $\sim 0.03$--$3$ yr over the $10^{-8}$--$10^{-5}$ bar region. In contrast, for the weakly irradiated cases ($0.1\,S_\oplus$), it decreases to the order of minutes, implying that vertical transport can dominate the upper-atmosphere composition in those cases. For the species dominating the opacity, $\mathrm{H_2O}$, $\mathrm{CO_2}$, $\mathrm{H_2}$, and $\mathrm{H_2S}$, the dominant loss pathways include direct photolysis of $\mathrm{H_2O}$, $\mathrm{CO_2}$, and $\mathrm{H_2S}$, $\mathrm{H_2O + h\nu \rightarrow OH + H}$, $\mathrm{CO_2 + h\nu \rightarrow CO + O}$, $\mathrm{H_2S + h\nu \rightarrow SH + H}$, as well as indirect chemical destruction of $\mathrm{H_2}$ through radical reactions such as $\mathrm{H_2 + OH \rightarrow H_2O + H}$. 

The corresponding photolysis timescale is $\tau_{\rm {\rm photo},i} = 1/J_{\rm i}$, where $J_{\rm i}$ is the total photolysis rate of species i. Over the $10^{-8}$--$10^{-5}$ bar region, $\tau_{\rm photo}$ decreases strongly with instellation, especially in the M-dwarf cases. For example, at $1000\,S_\oplus$ in the M-dwarf cases, median $\tau_{\rm photo}$ values are $\sim 1$--$10^4\,\mathrm{s}$ for $\mathrm{H_2O}$, $\sim 10^2$--$10^7\,\mathrm{s}$ for $\mathrm{CO_2}$, and $\sim 1$--$10^2\,\mathrm{s}$ for $\mathrm{H_2S}$. Among these species, $\mathrm{H_2S}$ exhibits the shortest loss timescales overall, implying extremely rapid sulfur processing in the observable atmosphere. By contrast, $\mathrm{H_2}$ generally has a comparatively longer $\tau_{\rm photo}$, particularly in the M-dwarf cases: at $1000\,S_\oplus$, median $\tau_{\rm photo}$ ranges from $\sim 10^3\,\mathrm{s}$ to effectively infinite values in the reduced cases. This shows that $\mathrm{H_2}$ depletion is controlled primarily by secondary radical chemistry rather than by direct photolysis. In contrast, $\mathrm{CO_2}$ remains photolytically longer-lived in the reduced and weakly irradiated cases, with median $\tau_{\rm photo}$ from days to centuries or longer depending on redox state and instellation. Comparing these values with $\tau_{\rm mix}$, which is typically $\sim 0.02$--$14$ yr at $1000\,S_\oplus$ but falls to the order of minutes at $0.1\,S_\oplus$, shows that the upper atmospheres of the highest-instellation cases are largely photochemically controlled, whereas at low instellation vertical mixing can more effectively preserve the compositional signature of the deeper outgassed reservoir. A detailed plot with the timescales over multiple species mentioned in this work, such as $\mathrm{H_2O, CO_2, H_2, CH_4, CO, SO_2, H_2S}$, and $\mathrm{SH}$, is shown in Appendix \ref{Asec: timescales}.

\subsection{Sulfur species}\label{sec:sulfur_species}

\begin{figure*}[h!]
    \centering
    \includegraphics[width=\textwidth]{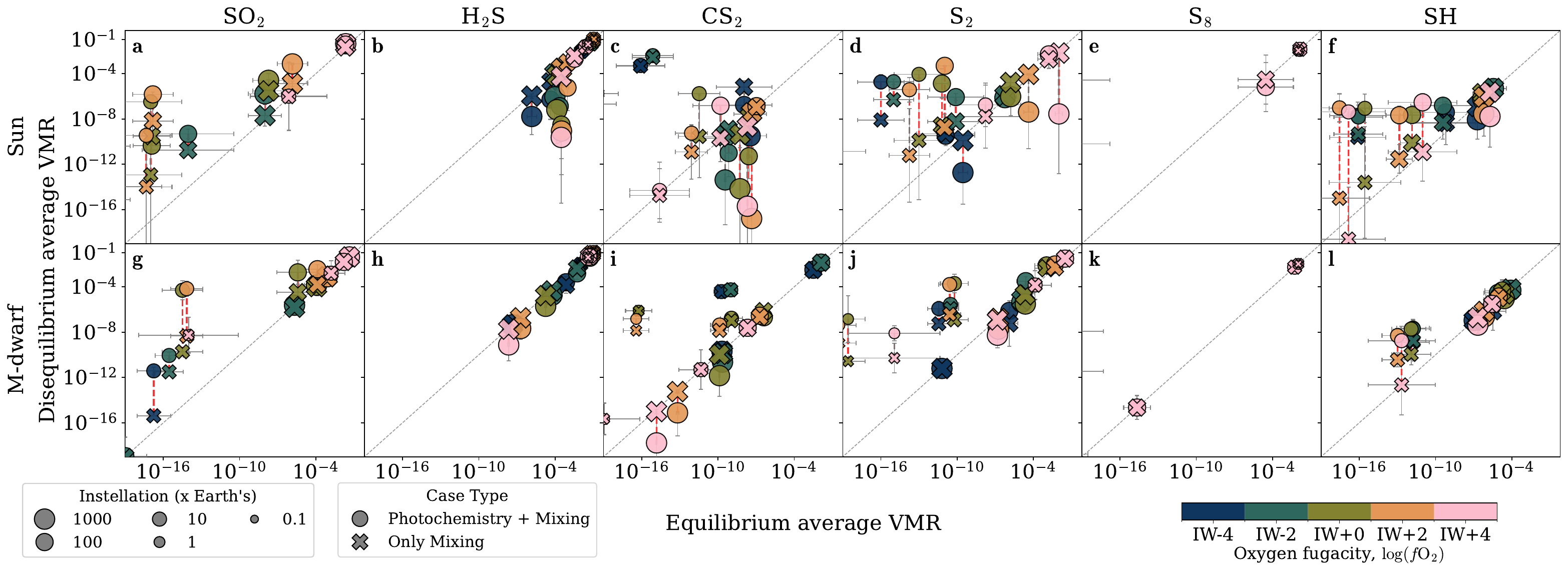}
    \caption{Comparison of the species abundances after the equilibrium and disequilibrium analysis for the Sun and M-dwarf cases. Each panel shows one important sulfur species ($\mathrm{SO_2, H_2S, CS_2, S_2, S_8}$, and SH), with the x-axis giving the equilibrium log-mean volume mixing ratio and the y-axis giving the corresponding disequilibrium value averaged over $10^{-4}$--$10^{-2}$ bar. The error bars represent the standard deviation divided by the number of points for the volume mixing ratios between these two pressure layers. The circles denote the result for mixing and photochemistry, while the X symbols denote the result with mixing only. The dashed red connectors link points from the same case at a fixed equilibrium abundance to distinguish the photochemical from the mixing results. The marker colour encodes the mantle redox state $\log (f\mathrm{O_2})$, the marker size encodes instellation ($0.1$--$1000\,S_\oplus$), and the dashed black diagonal marks the 1:1 agreement with equilibrium. Deviations from the diagonal quantify departures from thermochemical equilibrium, with the upper-left part corresponding to enhancement and the lower-right part corresponding to depletion.}
    \label{fig:sulfur_behaviour}
\end{figure*}

As discussed in Sect. \ref{sec: Introduction}, sulfur-bearing molecules provide a potentially powerful observational fingerprint of photochemical processing in rocky-planet atmospheres. In particular, $\mathrm{SO_2}$ has often been highlighted as a tracer of oxidising sulfur chemistry because it can be produced through the photochemical oxidation of reduced sulfur gases such as $\mathrm{H_2S}$, initiated by the photodissociation of $\mathrm{H_2O}$ and the subsequent production of OH and H radicals \citep{Zahnle_2016, Tsai_2023, Powell_2024}. However, the sulfur network is broader than $\mathrm{SO_2}$ alone. Since Sect. \ref{subsec: overall} showed that all mantle redox states maintain a significant abundance of outgassed $\mathrm{H_2S}$ in the bulk atmospheric composition, a range of sulfur-bearing species can be expected to form through equilibrium chemistry, mixing, and photochemical reprocessing. Figure \ref{fig:sulfur_behaviour} therefore compares the equilibrium and disequilibrium abundances of six important sulfur species: $\mathrm{SO_2}$, $\mathrm{H_2S}$, $\mathrm{CS_2}$, $\mathrm{S_2}$, $\mathrm{S_8}$, and SH, averaged over the photospheric region of the atmosphere ($10^{-4}$--$10^{-2}$ bar), which can be probed by transmission spectroscopy.

The behaviour of $\mathrm{SO_2}$ reflects both interior supply and atmospheric processing. Its equilibrium abundance increases strongly with mantle redox state, from $10^{-11}$ for IW-4 to $10^{-2}$ for IW+4 in the Sun cases (Fig. \ref{fig:sulfur_behaviour} \textbf{a}), showing that volcanic outgassing already favours $\mathrm{SO_2}$ in oxidised atmospheres. In addition, equilibrium chemistry in the hotter, highly irradiated cases shifts sulfur away from $\mathrm{H_2S}$ and towards more oxidised sulfur species. For example, reactions such as $\mathrm{S + H_2 \rightleftharpoons H_2S}$ become less favourable at high temperatures, reducing the abundance of $\mathrm{H_2S}$ (see decrease in average volume mixing ratio over all redox states in Fig. \ref{fig:sulfur_behaviour} \textbf{b}) and leaving more sulfur in atomic and radical form, while pathways such as $\mathrm{SO + H_2O \rightleftharpoons SO_2 + H_2}$ promote $\mathrm{SO_2}$ production when oxidising agents are available. When examining more oxidised scenarios, the atmospheric abundance of $\mathrm{H_2}$ decreases (as explained in Sect. \ref{subsec: overall}), which further shifts the sulfur chemistry towards $\mathrm{SO_2}$. In the disequilibrium calculations, $\mathrm{SO_2}$ is further enhanced at high instellation through photochemically mediated oxidation of reduced sulfur gases, with its average volume mixing ratio increasing by as much as 4 orders of magnitude for the intermediate-to-oxidised cases with $\geq\,10\,S_\oplus$ (see cases with dashed red lines in Fig. \ref{fig:sulfur_behaviour} \textbf{a} and \textbf{g}). Because $\mathrm{SO_2}$ is also a strong UV absorber and can be rapidly converted into sulfate aerosols in sufficiently oxidising atmospheres, it is both a tracer of ongoing sulfur cycling and a potentially important climate agent \citep{Hu_2013, Pitari_2016}.

By contrast, $\mathrm{H_2S}$ remains the dominant sulfur species in volcanic degassing from most mantles and shows a mostly consistent presence over different redox states. Its abundance remains high in equilibrium for reduced cases, but it is highly sensitive to disequilibrium processes, even at $10^{-4}$--$10^{-2}$ bar, because stellar UV photolysis rapidly breaks down $\mathrm{H_2S}$ into SH and other sulfur radicals, which act as intermediates for further processing. In strongly irradiated atmospheres ($\sim\,1000\,S_\oplus$), $\mathrm{H_2S}$ is efficiently converted into SH, atomic sulfur, $\mathrm{S_2}$, and ultimately more oxidised sulfur species such as $\mathrm{SO_2}$. This behaviour is strongest in the Sun cases (Fig. \ref{fig:sulfur_behaviour} \textbf{b}), where the disequilibrium abundance of $\mathrm{H_2S}$ departs most strongly from the equilibrium expectation, but it is still occurring for the M-dwarf cases (Fig. \ref{fig:sulfur_behaviour} \textbf{h}). The depletion of $\mathrm{H_2S}$ therefore marks the onset of active sulfur photochemistry and provides the feedstock for the rest of the sulfur network.

\begin{figure*}[h!]
    \centering
    \includegraphics[width=\textwidth]{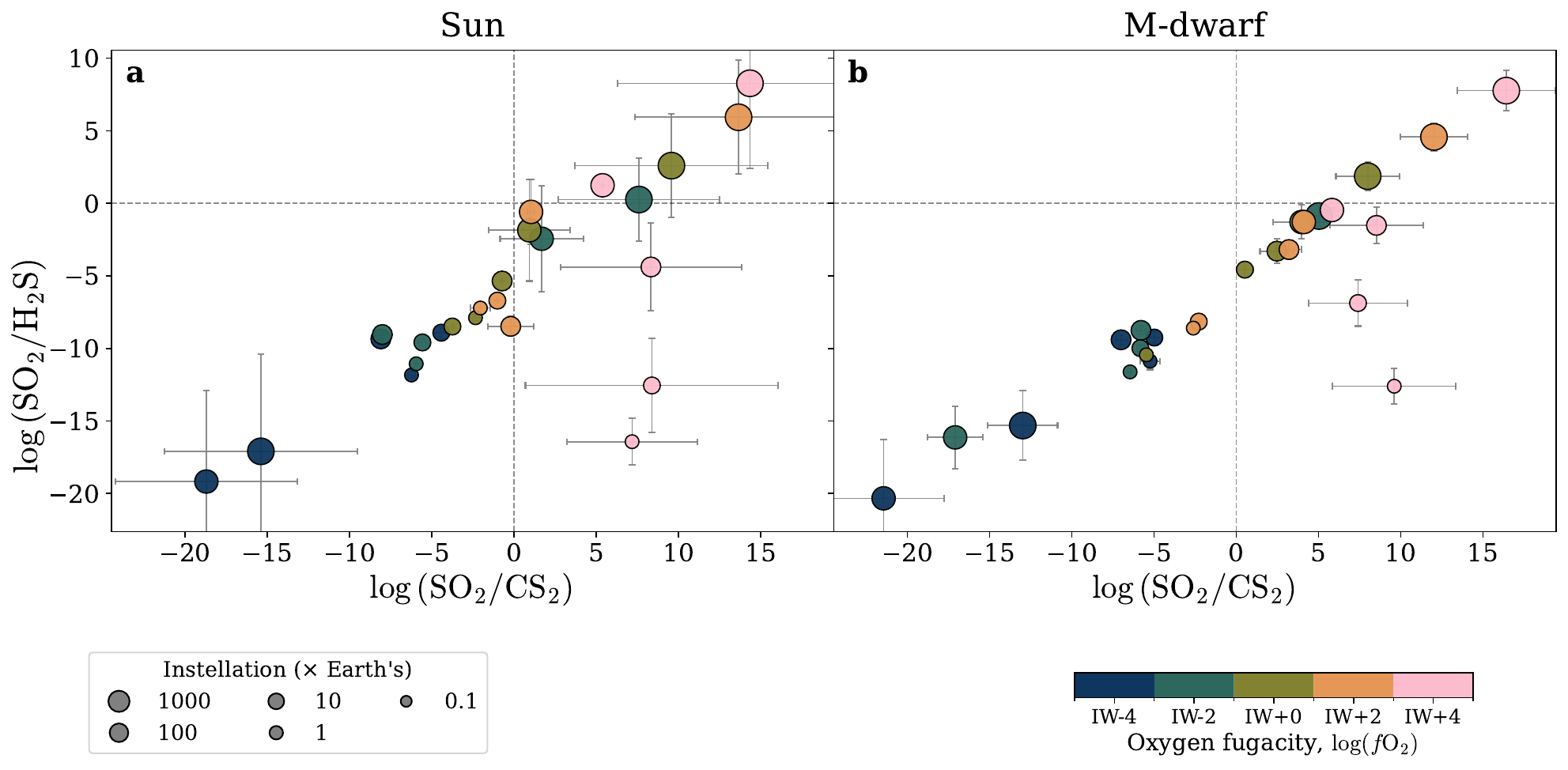}
    \caption{Logarithm of the ratio between the average volume mixing ratio of $\mathrm{SO_2}$ over $\mathrm{H_2S}$, as given by the disequilibrium calculation that includes photochemistry for the pressure layers between $10^{-2}$ and $10^{-4}$ bar, with respect to the ratio between $\mathrm{SO_2}$ over $\mathrm{CS_2}$. The error bars represent the standard deviation divided by the number of points for the VMRs between these two pressure layers. The marker colour encodes mantle redox state $\log (f\mathrm{O_2})$, and the marker size encodes instellation ($0.1$--$1000\,S_\oplus$).}
    \label{fig:abundance_ratios}
\end{figure*}

\begin{table*}[h!]
\centering
\caption{Approximate ranges of the average ratio between the disequilibrium abundances of $\mathrm{SO_2, CS_2}$, and $\mathrm{H_2S}$ in the photospheric region corresponding to different instellation fluxes and mantle redox states (based on Fig. \ref{fig:abundance_ratios}).}
\begin{tabular}{ccccc}
\hline \hline
Regime & $\log(\mathrm{SO_2/CS_2})$ & $\log(\mathrm{SO_2/H_2S})$ & Instellation & Likely mantle redox state \\\hline
1 & $[0,\,20]$ & $[0,\,10]$ & $1000\,S_\oplus$ & IW+0 to IW+4 \\
2 & $[0,\,15]$ & $[-20,\,0]$ & $0.1$--$100\,S_\oplus$ & IW+4 \\
3 & $[-5,\,5]$ & $[-12,\,-5]$ & $0.1$--$10\,S_\oplus$ & IW+0 to IW+4 \\
4 & $[-10,\,-5]$ & $[-12,\,-5]$ & $0.1$--$10\,S_\oplus$ & IW$-$2 to IW$-$4 \\
5 & $[-20,\,-10]$ & $[-20,\,-15]$ & $100$--$1000\,S_\oplus$ & IW$-$2 to IW$-$4 \\
\hline
\end{tabular}
\label{tab:sulfur_ratio_diagnostics}
\end{table*}

The more-reduced sulfur species $\mathrm{CS_2}$ and $\mathrm{S_2}$ are most prominent in atmospheres that are sulfur-rich (such as for IW+4 in Fig. \ref{fig:sulfur_behaviour} \textbf{d}) and chemically reducing. $\mathrm{CS_2}$ is generally a trace species, but Fig. \ref{fig:sulfur_behaviour} \textbf{c} and \textbf{i} show that it can be enhanced under intermediate-to-reduced conditions, particularly when high temperatures favour gas-phase sulfur-carbon reactions. However, very strong irradiation still causes photodissociation with increasing redox state, similar to $\mathrm{H_2S}$. $\mathrm{CS_2}$ is also UV-sensitive and can be rapidly destroyed when stellar irradiation is strong, which helps explain the broad departures from equilibrium seen in the disequilibrium calculations. Its abundance remains much lower than that of $\mathrm{H_2S}$ or $\mathrm{SO_2}$, but it is still useful as a marker of strongly reducing (such as for IW-4 in Fig. \ref{fig:sulfur_behaviour} \textbf{i}), carbon-rich volcanic environments. For reduced cases at 10 and $100\,S_\oplus$, mixing can help enhance its abundance in the photospheric region as well. $\mathrm{S_2}$ (Fig. \ref{fig:sulfur_behaviour} \textbf{d} and \textbf{j}) behaves as a short-lived intermediate: it is produced through photochemical processing of $\mathrm{H_2S}$ and other sulfur-bearing precursors, becomes enhanced in atmospheres with active sulfur photochemistry, and is especially favoured in intermediate-to-reducing redox states where enough reduced sulfur is present to feed the sulfur chain. Because $\mathrm{S_2}$ is highly reactive and a strong UV absorber, its atmospheric abundance is transient and signals ongoing photochemical reprocessing rather than simple outgassing alone.

Although it remains a minor gas-phase constituent in most of our cases, $\mathrm{S_8}$ appearance in Fig. \ref{fig:sulfur_behaviour} \textbf{e} and \textbf{k} is important because it traces the polymerisation of sulfur into species that can act as aerosol precursors. At sufficiently low temperatures, $\mathrm{S_8}$ is expected to condense and contribute to photochemical hazes or sulfur clouds, which can scatter efficiently at UV and visible wavelengths and produce an anti-greenhouse cooling effect. The presence of $\mathrm{S_8}$ is therefore less important as a bulk gas-phase absorber than as a signpost of a photochemically processed sulfur atmosphere capable of forming stable aerosol layers. In our results, it appears as a consistent element of very oxidised atmospheres at a volume mixing ratio of around $10^{-2}$. With increasing irradiation that can halt the production of $\mathrm{S_8}$, its value drops to $10^{-5}$ for the Sun (Fig. \ref{fig:sulfur_behaviour} \textbf{e}), and $10^{-15}$ for the M dwarf (Fig. \ref{fig:sulfur_behaviour} \textbf{k}).

Finally, SH is the most reactive sulfur species shown in Fig. \ref{fig:sulfur_behaviour} \textbf{f} and \textbf{l}. It is produced primarily through the dissociation of $\mathrm{H_2S}$ ($\mathrm{H_2S \rightarrow SH + H}$), and acts as a key radical intermediate linking reduced sulfur to more oxidised products. Its abundance is strongly enhanced in the disequilibrium calculations for the Sun, particularly at high instellation (see increasing abundance in the upper-left part of Fig. \ref{fig:sulfur_behaviour} \textbf{f}), consistent with its role as a short-lived intermediate in the conversion of $\mathrm{H_2S}$ into $\mathrm{S_2}$, $\mathrm{SO}$, and ultimately $\mathrm{SO_2}$. Because SH is extremely reactive, it does not build up to dominate the bulk sulfur budget. Instead, its presence traces the rate of sulfur cycling and the strength of atmospheric disequilibrium, similar to $\mathrm{S_2}$. It can maintain a more consistent abundance for the M-dwarf cases in Fig. \ref{fig:sulfur_behaviour} \textbf{l}. 

Figure \ref{fig:abundance_ratios} shows the logarithm of the ratio between the average volume mixing ratio of $\mathrm{SO_2}$ over $\mathrm{H_2S}$, as given by the disequilibrium calculation that includes photochemistry for the pressure layers between $10^{-2}$ and $10^{-4}$ bar, with respect to the ratio between $\mathrm{SO_2}$ over $\mathrm{CS_2}$. These three species are observable within the JWST wavelength range \citep{Polman_2023, Tsai_2023, Dai_2026} and illustrate the effects of photochemistry as a function of the mantle redox state and instellation flux. Both Fig. \ref{fig:abundance_ratios} \textbf{a} and \textbf{b} can be roughly separated into 5 regimes based on the comparison of the abundance ratios between the 3 species in the photospheric region (shown in Table \ref{tab:sulfur_ratio_diagnostics}):  high instellation ($1000\,S_\oplus$) and intermediate-to-oxidised (IW+0 to IW+4), low instellation ($0.1$--$100\,S_\oplus$) and most oxidised (IW+4), low instellation ($0.1$--$10\,S_\oplus$) and intermediate-to-oxidised (IW+0 to IW+4), low instellation ($0.1$--$10\,S_\oplus$) and reduced (IW$-$2 to IW$-$4), and high instellation ($100$--$1000\,S_\oplus$) and reduced (IW$-$2 to IW$-$4). For the low instellation middle region, reduced cases cluster more in Fig. \ref{fig:abundance_ratios} \textbf{b} compared to \textbf{a}, and provide a cleaner separation from the oxidised cases. As mentioned earlier, $\mathrm{SO_2}$ is enhanced in intermediate and oxidised cases, and increases the abundance ratios to the positive regions of Fig. \ref{fig:abundance_ratios}. $\mathrm{CS_2}$ is a species that is more prominent in reduced cases, similar to $\mathrm{H_2S}$. The abundance ratios of these observable species highlight that sulfur photochemistry can also provide a way to trace the mantle redox states of rocky planets in a more robust, instrument-independent way.

Overall, Figs. \ref{fig:sulfur_behaviour} and \ref{fig:abundance_ratios} show that sulfur chemistry in these atmospheres is controlled by a combination of mantle redox state and stellar irradiation. Reduced mantles favour volcanic outgassing of $\mathrm{H_2S}$, and the formation of reduced sulfur intermediates such as $\mathrm{SH}$ and $\mathrm{S_2}$, whereas oxidised mantles favour $\mathrm{SO_2}$-rich atmospheres. Mixing is most important in the low-instellation cases, where chemical timescales become long enough for vertical quenching to preserve deeper atmospheric abundances. Increasing instellation enhances photochemical processing, depleting $\mathrm{H_2S}$, and driving sulfur towards oxidised or polymerised end products. Photochemical processing dominates throughout the photospheric region, so the potentially observable abundance of sulfur species, such as $\mathrm{SO_2, H_2S}$, and $\mathrm{CS_2}$, reflects not only volcanic outgassing but also the extent to which stellar irradiation has reprocessed the sulfur network.

\subsection{Synthetic spectra}

\begin{figure}[t]
    \centering    \includegraphics[width=\columnwidth]{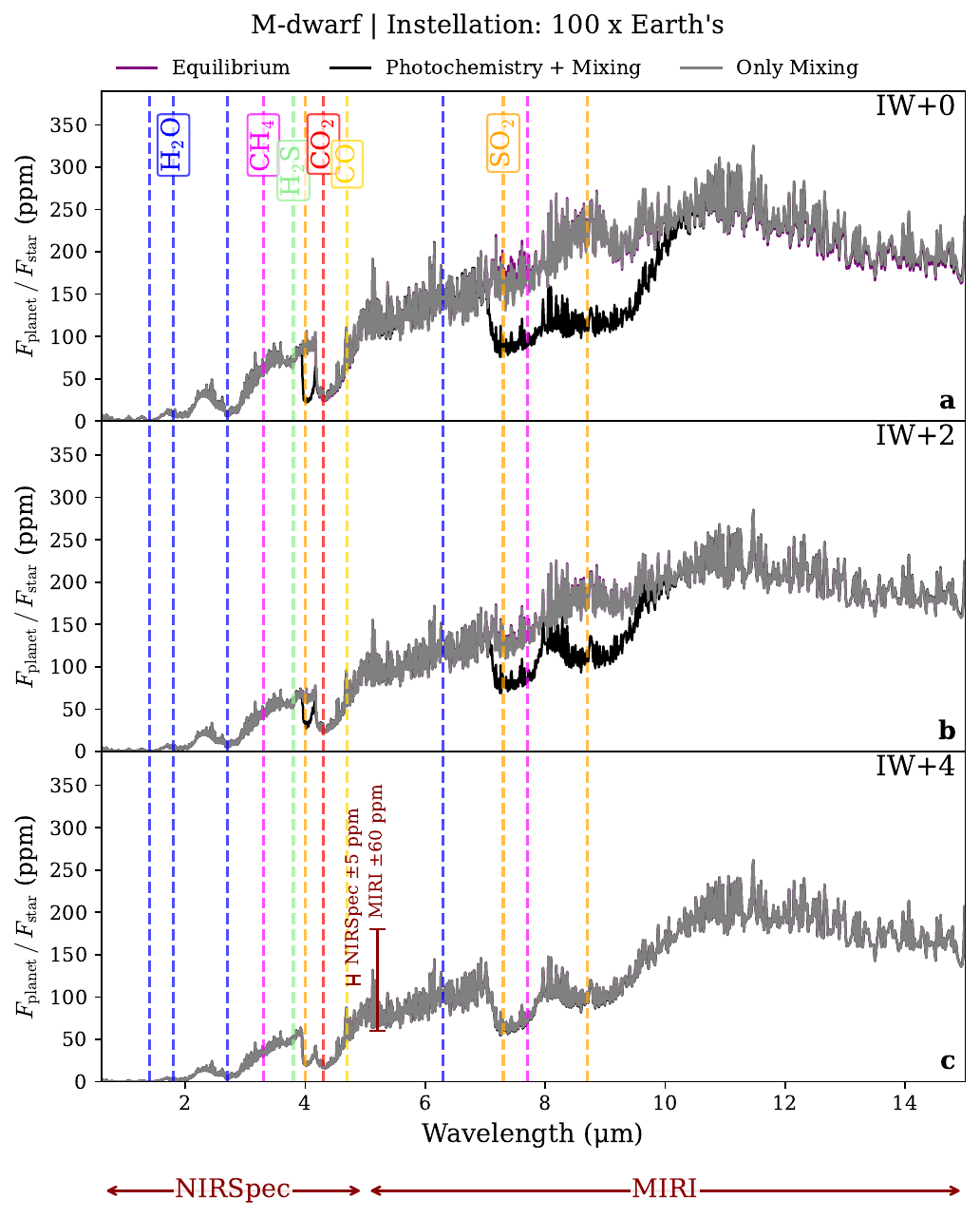}
    \caption{Equilibrium and disequilibrium synthetic emission spectra generated with \texttt{AGNI} for three M-dwarf cases with $100\,S_\oplus$. The cases shown correspond to the intermediate and oxidised mantle redox states with IW+0, IW+2, and IW+4. The y-axis shows the ratio between the planet's flux and the stellar flux, whereas the x-axis is the wavelength range for JWST's NIRSpec and MIRI (up to $15\,\mu$m). Notable absorption features are indicated with the coloured vertical lines, and the average uncertainty in the data points are highlighted in red.}
    \label{fig:spectrum_selected}
\end{figure}

To inspect for observational imprints from photochemical processes, synthetic emission spectra generated with \texttt{AGNI} are shown in Fig. \ref{fig:spectrum_selected} for $100\,S_\oplus$ over the three M-dwarf mantle redox states of IW+0 (\textbf{a}), IW+2 (\textbf{b}), and IW+4 (\textbf{c}). The wavelength range selected for visualisation corresponds to JWST's NIRSpec and MIRI instruments (up to $15\,\mu$m). Notable spectral features can be seen for various species, such as $\mathrm{H_2O}$ and $\mathrm{CO_2}$, which dominate the bulk atmospheric composition. Additionally, the near overlap of the equilibrium and mixing-only spectra in Fig. \ref{fig:spectrum_selected} \textbf{a-c} indicates that vertical quenching alone does not substantially modify the dominant opacity sources in the photospheric region for these cases, and that the main observable departures from equilibrium arise from photochemical production of $\mathrm{SO_2}$ instead. Its presence reduces the planet–star flux contrast at $4\,\mu$m from 80 to 25 ppm (see black line in Fig. \ref{fig:spectrum_selected} \textbf{a}). Two more absorption features can be seen in MIRI's wavelength range with notable flux ratio differences:  $7.3\,\mu$m (175 to 75 ppm), and $8.7\,\mu$m (225 to 125 ppm). More specifically, $\mathrm{SO_2}$'s NIRSpec feature at $4\,\mu$m acquires a signature of $\sim$60 ppm for IW+0, $\sim$40 ppm for IW+2, and returns to the outgassed value of $\sim$30 ppm for IW+4. The MIRI features at 7.3 and $8.7\,\mu$m acquire a signature of $\sim$100 ppm for IW+0, $\sim$70 ppm for IW+2, and return to the outgassed value of $\sim$50 ppm for IW+4. The reduction in planet-star flux contrast is consistent with enhanced $\mathrm{SO_2}$ opacity, shifting the wavelength-dependent photosphere to higher and cooler atmospheric layers, thereby lowering the emergent thermal flux within these bands \citep{Seager_2010}. 

The effect on the spectrum decreases for more oxidising redox states, with IW+4 showing $\mathrm{SO_2}$ features that are already present in the bulk composition rather than being produced primarily by photochemistry (Fig. \ref{fig:spectrum_selected} \textbf{c}). This weakening is also consistent with the increase in mean molecular weight $\mu$ across these cases, from $\mu \approx 14.4$ at IW+0 to $\mu \approx 22.0$ at IW+2 and $\mu \approx 25.2$ at IW+4, which lowers the atmospheric scale height and may further reduce the spectral contrast \citep{Madhusudhan_2019}, in addition to the reduced relative importance of photochemically produced $\mathrm{SO_2}$ itself. This result demonstrates that photochemistry can produce observable spectral signatures even when the bulk atmospheric composition remains largely unchanged, implying that atmospheric retrievals that neglect disequilibrium sulfur chemistry may misattribute these features to equilibrium abundances or to different underlying elemental inventories \citep{Hu_2012, Changeat_2019, Kawashima_2021, Nicholls_2026}.

\section{Discussion}\label{sec: Discussion}
\subsection{Bulk atmospheric composition and dominant photochemistry regimes}
A key result of this work is that photochemistry does not significantly alter the bulk composition of secondary, outgassed atmospheres. These atmospheres remain primarily determined by mantle redox state and outgassing history, highlighting the role of interior-atmosphere processes in determining the climates of low-mass exoplanets \citep{Lichtenberg_2025}. Across all simulations, the dominant species identified in equilibrium are preserved even when mixing and photochemistry are included. This indicates that, for Earth-like rocky planets with 1000-bar secondary atmospheres, atmospheric chemistry acts as a secondary modifier to the outgassing behaviour. The majority of atmospheric mass resides at high pressures deep in the atmosphere, where temperatures are higher and chemical reaction rates are sufficiently rapid to favour thermochemical equilibrium \citep{Zahnle_2009, Moses_2014, Drummond_2018}. In contrast, photochemical processes operate most efficiently in the low-pressure upper atmosphere, which contributes negligibly to the total column density. As a result, even strong photodissociation does not propagate deeply enough to modify the bulk inventory. This finding reinforces the framework proposed by \citet{Lichtenberg_2021} and \citet{Liggins_2022}, where planetary classification into reduced, oxidised, and intermediate atmospheres is fundamentally controlled by interior redox evolution, rather than stellar irradiation. \citet{Lichtenberg_2021} employ a fully molten mantle and \citet{Liggins_2022} a volcanic source over a solid mantle, leading to different mole fraction predictions but similar classification.

Although photochemistry is subdominant for bulk composition, it becomes the dominant disequilibrium process in the upper atmosphere under sufficiently strong irradiation. We can identify two regimes: one for low instellation fluxes up to $10\,S_\oplus$ and one for high instellation fluxes up to $1000\,S_\oplus$. For the first case, mixing dominates over photochemistry, with vertical transport redistributing species such as $\mathrm{H_2}$, which often enhances its abundance in the upper atmosphere, and $\mathrm{CO_2}$, which gets transported into the deeper layers, by several orders of magnitude relative to equilibrium. This photochemical behaviour is supported by previous work, as described by \citet{Linsky_2014}. Photochemistry dominates in the second case. Strong XUV irradiation drives extensive photodissociation, producing radicals, with special enhancement of H, and significantly depleting stable molecules such as $\mathrm{H_2}$ and $\mathrm{CO_2}$ in the upper atmosphere. The transition between these regimes is controlled by the competition between photochemical and mixing timescales. At high irradiation, photolysis rates exceed transport rates, leading to large departures from equilibrium. This is particularly enhanced in M-dwarf environments, where elevated XUV flux increases photodissociation efficiency despite similar instellation. The competition between the disequilibrium processes shapes the final abundances, with hotter planets having quick enough reaction rates that overtake vertical transport effects, as described in \citet{Moses_2014}.

\citet{Drant_2025} explored the possibility of using the observational abundances of $\mathrm{CO_2}$ and CO to infer the interior oxygen fugacity $f\mathrm{O}_2$. However, enhancement of $\mathrm{CO_2}$ driven by $\mathrm{H_2O}$ photolysis, which leads to oxidation of CO, introduces bias in the predicted mantle redox state. In our work, $\mathrm{CO_2}$ and CO dominate the bulk composition in oxidised and intermediate cases, respectively, which agrees with their resulting atmospheres. However, these species experience only strong photodissociation for higher instellation fluxes, with mixing smoothing out their abundance under weaker irradiation. $\mathrm{SO_2}$ constitutes a stronger tracer of mantle redox state, due to the more prominent sulfur photochemical network and its notable signatures in the JWST wavelength range.

\subsection{Observational signatures of $\mathrm{SO_2}$}
One of the most significant outcomes is the behaviour of $\mathrm{SO_2}$ as a diagnostic of photochemical activity. Three regimes can be isolated: low instellation ($0.1\,S_\oplus$), intermediate ($1$--$100\,S_\oplus$), and extreme instellation ($1000\,S_\oplus$). For the low instellation regime, $\mathrm{SO_2}$ follows equilibrium chemistry and remains primarily controlled by mantle redox state. \citet{Liggins_2022} show that more oxidised cases result in outgassing of $\mathrm{SO_2}$, and Fig. \ref{fig:cdrs} shows the presence of $\mathrm{SO_2}$ in the bulk composition for the high instellation plus IW+4 case, agreeing with their conclusions. For the intermediate irradiation regime, photochemical production becomes highly efficient, especially at redox states comparable to the iron-w\"ustite buffer. This leads to orders-of-magnitude enhancements in the $\mathrm{SO_2}$ abundance in the photosphere, resulting in absorption signatures of 50--100 ppm at 4, 7.3 and $8.7\,\mu$m that are potentially accessible to JWST instruments. 

The precision in data collected with JWST's NIRSpec and MIRI have a wide range of values depending on the exoplanet, the number of transits or eclipses, the data reduction methods, and observational circumstances. \citet{Monaghan_2026} collected the uncertainties in multiple rocky planet eclipse observations with JWST to explore the effects of the host star and the planet's orbit. The planet-to-star contrast error bars for MIRI (F1500W and F1280W) vary from 26 ppm for GJ~3929\,b \citep{Xue_2025} to 128 ppm for TRAPPIST\,1--c \citep{Connors_2025}. Across all listed eclipse observations, the average error is 60 ppm, which is highlighted in Fig. \ref{fig:spectrum_selected}. For NIRSpec (G395H NRS1 and NRS2), the error bars vary from 3 ppm for TOI--561\,b \citep{Teske_2025} to 7 ppm for TOI--1685\,b \citep{Luque_2025}. Across all eclipse observations, the average error is 5 ppm, which is also highlighted in Fig. \ref{fig:spectrum_selected}. These uncertainties represent the state of rocky planet eclipse observations and illustrate the potential detectability of signatures such as $\mathrm{SO_2}$'s. Moreover, a feature strength metric could make the detection plausibility more robust \citep{Mansfield_2021}. Overall, the signatures at 4, 7.3, and $8.7\,\mu$m, and their connection to the mantle redox state are potentially detectable with JWST, motivating targeted observational campaigns.

For the extreme irradiation case, $\mathrm{SO_2}$ is photodissociated in the upper atmosphere, limiting its abundance at the lowest pressures. The peak enhancement at intermediate redox states reflects a balance between the availability of $\mathrm{H_2S}$ as the prominent sulfur source and the availability of oxidising radicals from $\mathrm{H_2O}$ photolysis. This sensitive behaviour suggests that $\mathrm{SO_2}$ detections cannot be interpreted purely as indicators of oxidised outgassing, but instead require accounting for photochemical pathways \citep{Tsai_2023, Nicholls_2026}. Since $\mathrm{SO_2}$ can be produced photochemically in the intermediate redox state or it can be outgassed in a more oxidised state, it is not a unique tracer of redox state by itself. The synthetic spectra demonstrate that photochemistry can produce observable signatures even when bulk composition is unchanged. Photochemistry can bias atmospheric retrievals. When excluded from the models, the retrieved abundances may incorrectly attribute features to equilibrium chemistry or different elemental compositions. Previous work has explored disequilibrium processes affecting exoplanet spectra and their detectable imprints, including photochemical kinetics in \citet{Moses_2014}, time-dependent reactions for hot Jupiters in \citet{Hobbs_2019}, and even considering 3D effects in \citet{Arora_2026}.

\subsection{Photochemical hazes}
An additional implication of the disequilibrium chemistry explored here is the potential formation of photochemical hazes, which may provide an important feedback between atmospheric chemistry, climate, and long-term planetary evolution \citep{Trainer_2006, Gao_2017}. Several species that appear in our simulations are commonly associated with haze formation pathways, including benzene and other higher-order hydrocarbons in reduced atmospheres, as well as sulfur-chain species and allotropes such as $\mathrm{S_2}$ and $\mathrm{S_8}$. Our comparison in Appendix \ref{Asec: benzene} with the species included in \texttt{FastChem} demonstrates that benzene can become abundant in the upper atmosphere of strongly reduced cases, indicating that carbon can be diverted from simple molecules such as $\mathrm{CH_4}$ into more complex hydrocarbons that act as haze precursors \citep{Arney_2018, He_2020}. Likewise, the disequilibrium sulfur chemistry discussed in Sect. \ref{sec:sulfur_species} shows that photolysis of $\mathrm{H_2S}$ efficiently feeds sulfur radicals and polymerisation pathways, making sulfur aerosol formation plausible in irradiated atmospheres \citep{Zahnle_2016}. Although we do not model aerosol microphysics explicitly, these results suggest that haze production may be a natural consequence of the photochemical processing identified here. If present, such hazes could affect atmospheric structure in several ways. First, they would alter the observable spectra by introducing broad continuum opacity, muting molecular absorption bands, and in some cases reshaping the apparent strengths of diagnostic features such as $\mathrm{SO_2}$, $\mathrm{H_2O}$, and $\mathrm{CO_2}$ \citep{Kawashima_2018}. Second, hazes would modify the thermal structure through wavelength-dependent absorption and scattering. Sulfur aerosols are expected to scatter efficiently in the UV and visible and may therefore produce an anti-greenhouse effect, cooling the lower atmosphere while reducing the penetration depth of stellar radiation \citep{McKay_1991, Arney_2017}. Hydrocarbon hazes, by contrast, can also absorb efficiently at short wavelengths, potentially heating the upper atmosphere while shielding deeper layers from photolysis \citep{Hu_2012}. In both cases, the redistribution of radiative heating and cooling would feed back on the atmospheric pressure-temperature structure and therefore on the chemical reaction network itself, since equilibrium constants, photolysis rates, and vertical mixing efficiencies all depend on local temperature and density \citep{Arney_2017, Kawashima_2019}. These effects are particularly relevant for the long-term thermal evolution of rocky planets in coupled interior-atmosphere systems. In \texttt{PROTEUS}, the atmospheric structure regulates the radiative cooling of the magma ocean and hence the timescale over which the mantle solidifies and outgasses volatiles \citep{Hamano_2013, Lichtenberg_2021b}. 

If photochemical hazes increase the planetary albedo or reduce the efficiency of outgoing thermal radiation, they could either delay or accelerate cooling depending on their vertical distribution and optical properties. Sulfur-rich hazes could promote atmospheric cooling through enhanced scattering, whereas strongly absorbing hydrocarbon hazes could contribute to upper-atmosphere heating and modify the vertical temperature gradient \citep{Hu_2012, Gao_2017, Arney_2017}. In either case, the presence of aerosols would alter the balance between absorbed stellar flux and thermal emission, and could therefore shift the timing of mantle crystallisation, volatile release, and the transition between magma-ocean and solid-mantle evolutionary pathways \citep{McKay_1991, Hamano_2013, Arney_2017}. 

Haze formation may further affect chemistry by attenuating UV photons, suppressing photolysis in deeper layers, and opening or closing reaction pathways through heterogeneous surface chemistry on particle grains \citep{Gao_2017, Lavvas_2017, Kawashima_2018}. A full assessment of their impact requires coupling the present chemistry calculations to aerosol microphysics and radiative transfer, which is beyond the scope of this work \citep{Gao_2017, Arney_2017, Kawashima_2018}. Nevertheless, our results already show that photochemistry can generate the precursor molecules needed for hazes, such as $\mathrm{S_8}$, implying that future models of rocky exoplanet atmospheres should treat aerosols as a potentially important link between atmospheric disequilibrium, climate, and interior evolution \citep{Zahnle_2016, Gao_2017}.

\subsection{Limitations, improvements, and suggested observationally accessible planets}
Several simplifying assumptions have been made in this work and should be addressed in future studies. No atmospheric escape was implemented in the coupling between the atmosphere and stellar irradiation, even though our atmospheres show a significant presence of atomic H in the upper layers. Strong photodissociation produces light species such as H that are susceptible to escape and can change the overall atmospheric composition and long-term evolution \citep{Zahnle_1986, Gregory_2023, Cangi_2023}.  In our work, we modelled the magma ocean phase of rocky planets and highlight atmospheric results following its termination. After billions of years, the escape-prone H-bearing species could change the connection between sulfur species and mantle redox state, especially for warm planets around M-dwarf stars \citep{Krissansen_Totton_2024}. However, hydrogen replenishment remains possible and can help maintain the magma-ocean atmospheric composition presented in our work \citep{Hu_2023, Arora_2026b}. 

\texttt{PROTEUS} is a 1D modelling framework with mixing-length theory convection, which allows $K_{\rm zz}$ diffusion coefficients to be estimated from the balance of convective and radiative energy transport. However, this approach neglects larger-scale atmospheric dynamics such as gravity wave breaking, zonal heat redistribution, and the presence of cold traps \citep{Parmentier_2013, Drummond_2018, Drummond_2020, Tsai_2024m, Parmentier_2026, Liu_2026}. Lower instellation flux cases (1 and $0.1\,S_\oplus$) represent magma ocean planets in or near the habitable zone for which temperatures drop to low values (see Fig. \ref{fig:pt_profiles}), potentially allowing condensation to occur, which is not implemented in our work but could impact the chemistry calculations. The effects of condensation are discussed in Appendix \ref{Asec: condensation}. These additional processes may induce vertical chemical gradients that are reflected in JWST observations. Incorporating these effects into coupled climate-chemistry models should be considered, for a potentially more accurate understanding of the atmosphere-interior for diverse Earth-sized exoplanets.

A natural next step would be to implement \texttt{VULCAN}'s photochemical network and vertical transport approach over the entire evolution of the rocky planets explored in this work. \texttt{VULCAN} is called as a final post-process calculation to the final outgassed atmosphere after convergence is reached. As discussed in Sect. \ref{subsec: overall}, photochemistry and mixing do not affect the bulk atmospheric composition, but exploring disequilibrium effects over the planet's entire evolution could potentially allow disentanglement between the two pathways discussed in Sects. \ref{sec: Introduction} and \citet{Lichtenberg_2021} through possible observable signatures. This comes down to limitations from \texttt{VULCAN}, which determines the atmospheric composition over multiple steps and varies in calculation time from minutes to hours per simulation snapshot, depending on convergence conditions \citep{Hendrix_2023}. Unlike equilibrium chemistry calculations done with \texttt{FastChem} that depend only on the temperature, pressure, and elemental abundances \citep{Stock_2022}, \texttt{VULCAN} needs to calculate the effects from advection, eddy diffusion, and molecular diffusion on top of photochemical chemistry. Therefore, expanding the photochemical calculations over a Myr planetary evolution is a limitation of this work.

Finally, a limitation appears in the specific type of rocky planets explored in this work. Earth-like planets with the same initial planetary inventory of C-H-O-N-S volatiles were investigated, with C/H = 1.0, N/H = 0.5, S/H = 2.0, and a total amount of hydrogen equivalent to five Earth oceans. The initial inventory of volatiles affects the evolution of the planet through the interior reactions and the distribution of species in the mantle. This leads to a different outgassing abundance and an overall different atmospheric composition. \citet{Ortenzi_2020}, \citet{Massol_2023}, and \citet{Baumeister_2025} demonstrate the importance of such an effect. The analysis presented in this work could be expanded by extending the parameter space over the initial inventory of C, H, N, O, and S. In addition, extending this work to super-Earths and sub-Neptunes, which have massive atmospheres, could be useful. Processes such as gas accretion from the initial stellar nebula, contraction due to cooling, atmospheric escape, migration from snow lines, and the retention of volatile species affect the evolution of these populations and, when coupled with photochemistry, could lead to a diverse range of atmospheres \citep{Drazkowska_2023, Krijt_2022, Lichtenberg_2022, Tian_2024}. Extending the types of planets tested with the framework outlined in this work would allow for exploration of disequilibrium effects on a wider rocky planet population.

Suggestions for observationally accessible planets to test the effects of photochemistry on their atmospheric composition, lying in the 100 and $1000\,S_\oplus$ bins as discussed in this work, are K2--141\,b \citep{Malavolta_2018}, 55 Cnc e \citep{McArthur_2004}, TOI--431\,b \citep{Osborn_2021}, TOI--561\,b \citep{Lacedelli_2021}, and HD~3167\,b \citep{Vanderburg_2016}. The super-Earth K2--141\,b ($5.08\,M_\oplus$, $1.51\,R_\oplus$) orbiting on a short 6.7-day period around a K-type star provides a unique opportunity to explore the effects of photochemistry on a silicate-dominated atmosphere over a magma ocean, and the behaviour of volatile species in this cycle of rock-vapour chemistry \citep{Nguyen_2020,Zilinskas2025AA}. The highly irradiated 55 Cnc e ($7.99\,M_\oplus$, $1.87\,R_\oplus$) and its suspected magma ocean provide an opportunity for investigating the interactions between its interior and atmosphere, with previous work by \citet{Dash_2025} showing that the mantle redox state can be inferred spectroscopically. Using data from Spitzer, \citet{Monaghan2025AJ} report an eclipse depth of TOI--431\,b ($3.07\,M_\oplus$, $1.28\,R_\oplus$) between 4 and 5 microns, the same range as $\mathrm{SO_2}$'s NIRSpec signature, with a temperature of 1520\,K that disfavours the bare-rock scenario, and suggest the potential presence of a molecular absorption feature. JWST observations of the super-Earth TOI--561\,b orbiting a 10 Gyr-old iron-poor star ($2.02\,M_\oplus$, $1.40\,R_\oplus$) point to the presence of a thick volatile envelope rich in carbon over its magma ocean, and could allow the exploration of photochemical effects and the presence of potential species such as $\mathrm{CS_2}$ on an evolved older planet \citep{Peng_2024, Teske_2025}. HD~3167\,b ($4.84\,M_\oplus$, $1.60\,R_\oplus$) is an under-dense lava world that lies on the lower regime of equilibrium temperatures at 1786\,K \citep{Coy_2026}, and probing for photochemical effects could constrain its atmospheric composition. Finally, the rocky planet LHS 1478\,b ($2.27\,M_\oplus$, $1.17\,R_\oplus$) has an equilibrium temperature of 585\,K and receives 21 $S_\oplus$ \citep{Soto_2021, August2025AA}, placing it in the range where photochemical reprocessing dominates over mixing ($>10\,S_\oplus$), as highlighted in our work.

\section{Conclusions}\label{sec: Conclusions}
Our simulations of Earth-sized rocky exoplanets couple the \texttt{PROTEUS} planetary evolution framework to the \texttt{VULCAN} chemical kinetics model to investigate how photochemistry drives outgassed atmospheres out of equilibrium and modulates the coupling between their interior geochemistry and observable atmospheres. Planetary evolution pathways were simulated up to a state when the mantle melt fraction drops below 1\%, or a quasi-steady state of global energy balance is reached. We ran a simulation grid covering reduced to oxidised mantles and increasing instellation fluxes, assuming two different stellar hosts: the Sun and the exoplanet-host M-dwarf GJ 1132. Three post-processing analyses were then performed on each evolution simulation to probe disequilibrium processes: 1D thermochemical equilibrium, chemical kinetics with disequilibrium mixing, and disequilibrium mixing with photochemical kinetics. 

Our main conclusions can be summarised as follows:
\begin{itemize}
    \setlength{\itemsep}{5pt}
    \setlength{\parskip}{5pt}
    \item Bulk atmospheric compositions are controlled by outgassing, which transports volatiles from deep planetary interiors into their atmospheres, while dependent on mantle redox state. Atmospheric mixing and photochemistry do not substantially change the bulk atmospheric composition. Chemically reduced mantles outgas $\mathrm{H_2}$ and $\mathrm{CH_4}$ atmospheres, while intermediate cases outgas $\mathrm{CO}$, and more oxidised cases outgas $\mathrm{CO_2}$. Outgassing history is affected by the received irradiation and mantle melt fraction, which act as secondary drivers and can generate $\mathrm{H_2O}$-atmospheres.

    \item Stellar UV irradiation strongly impacts upper atmosphere compositions ($10^{-5}$ to $10^{-8}$ bar) by photodissociating molecular species and promoting radical formation, mostly H. Chemical species comprising the bulk atmosphere, $\mathrm{H_2}$ and $\mathrm{CO_2}$, are depleted in the upper atmosphere, which is correlated with mantle oxidation state. This trend is apparent for our 100 and $1000\,S_\oplus$ in M-dwarf cases but only for $1000\,S_\oplus$ for the solar host. Differences from the M-dwarf's UV radiation impact planets' photochemistry.

    \item Disequilibrium mixing causes compositional quenching for low instellation fluxes ($<10\,S_\oplus$) -- independent of mantle redox state -- acting to homogenise vertical gradients. Mixing can enhance the presence of $\mathrm{H_2}$ in the upper atmosphere by quenching its abundances at its surface-outgassed values.

    \item Sulfur chemistry is simultaneously controlled by mantle redox state, stellar irradiation, and vertical mixing. Outgassed $\mathrm{H_2S}$ is readily photodissociated, resulting in the formation of sulfur intermediates $\mathrm{S_2}$ and $\mathrm{SH}$. Oxidised mantles favour atmospheres rich in $\mathrm{SO_2}$, which at high instellations $\geq 100\,S_\oplus$ becomes photochemically enhanced, promoting sulfur allotropes $\mathrm{S_8}$. Observational comparison of sulfur species ($\mathrm{H_2S, SO_2}$ and $\mathrm{CS_2}$) abundances can trace mantle redox conditions.

    \item $\mathrm{SO_2}$ forms on planets with mantle redox states greater than or equal to the~IW~buffer for instellations comparable to 100 $S_\oplus$ through its photochemical production. Equilibrium outgassing yields abundant $\mathrm{SO_2}$ for redox states $\ge\mathrm{IW}$+2. Absorption by photochemical $\mathrm{SO_2}$ reduces the planet–star flux emission contrast at $4\,\mu$m (from 80 to 25 ppm) at redox states $\sim\mathrm{IW}$. MIRI-accessible features also show notable absorption due to $\mathrm{SO_2}$ production: $7.3\,\mu$m (from 175 to 75 ppm) and $8.7\,\mu$m (from 225 to 125 ppm). $\mathrm{SO_2}$'s NIRSpec-accessible feature at $4\,\mu$m generates an absorption depth of $\sim$60 ppm for IW+0 redox conditions, $\sim$40 ppm for IW+2, and $\sim$30 ppm for IW+4. The MIRI features at 7.3 and $8.7\,\mu$m generate a $\sim$100 ppm deep feature for IW+0 conditions, $\sim$70 ppm for IW+2, and resolve to the equilibrium-outgassed conditions of $\sim$50 ppm at IW+4. Robust characterisation of sulfur chemistry, with JWST, may trace planets' deep interior geochemistry.

\end{itemize}

Planetary interior redox states and subsequent volatile outgassing dictate the bulk atmospheric composition of Earth-sized planets. However, the disequilibrium effects of photochemistry and mixing modulate their upper atmospheres.  Spectrally active species (e.g. $\mathrm{SO_2}$) can alter the atmospheric structure and leave observable imprints. Coupled interior–atmosphere outgassing and photochemical processes are both leading-order effects to be considered when interpreting observations from JWST and the ESA space mission \textit{Ariel}.

\begin{acknowledgements}

CRediT author statements. Conceptualisation: TL, SMT, HN.

Methodology: HN, SMT, TL, IP.

Software: HN, SMT, TL.

Validation: IP, TL, SMT, HN.

Formal analysis: IP.

Investigation: IP.

Resources: TL.

Data Curation: IP.

Writing -- Original Draft: IP.

Writing -- Review \& Editing: IP, TL, SMT, HN.

Visualisation: IP.

Supervision: TL, SMT, HN.

Project Administration: TL.

Funding Acquisition: TL.

This research was supported by the Branco Weiss Foundation, the European Research Council (ERC) under the European Union's Horizon Europe research and innovation programme (MagmaWorlds, 101219807), the Alfred P. Sloan Foundation (AEThER, G-2025-25284), NASA’s Nexus for Exoplanet System Science research coordination network (Alien Earths, 80NSSC21K0593), and the NWO NWA-ORC PRELIFE Consortium (NWA.1630.23.013). S-.M. T. is supported by the National Science and Technology Council (grants 114-2112-M-001-065-MY3) and an Academia Sinica Career Development Award (AS-CDA-115-M03). HN acknowledges support from STFC grant UKRI1184. We thank the Center for Information Technology of the University of Groningen for their support and for providing access to the H\'abr\'ok high-performance computing cluster. We thank the anonymous referee for their positive and helpful review. Colourblind-friendly colourmaps were taken from \citet{Crameri_2020}.
\end{acknowledgements}

\bibliographystyle{aa}
\bibliography{references.bib}

\begin{appendix}
\nolinenumbers
\section{Relevant timescales}\label{Asec: timescales}
To interpret the upper atmosphere behaviour described in Sect. \ref{subsec: top of atm}, two characteristic timescales at each pressure level are computed: mixing and photolysis. Comparing these quantities provides a compact diagnostic of whether transport or local photochemistry controls the abundance of each species at a given pressure level. When $\tau_{\rm photo}$ becomes shorter than $\tau_{\rm mix}$, the species is processed locally faster than it can be vertically redistributed, whereas the opposite regime indicates stronger transport control. 

\begin{figure}[t]
    \centering
    \includegraphics[width=\columnwidth]{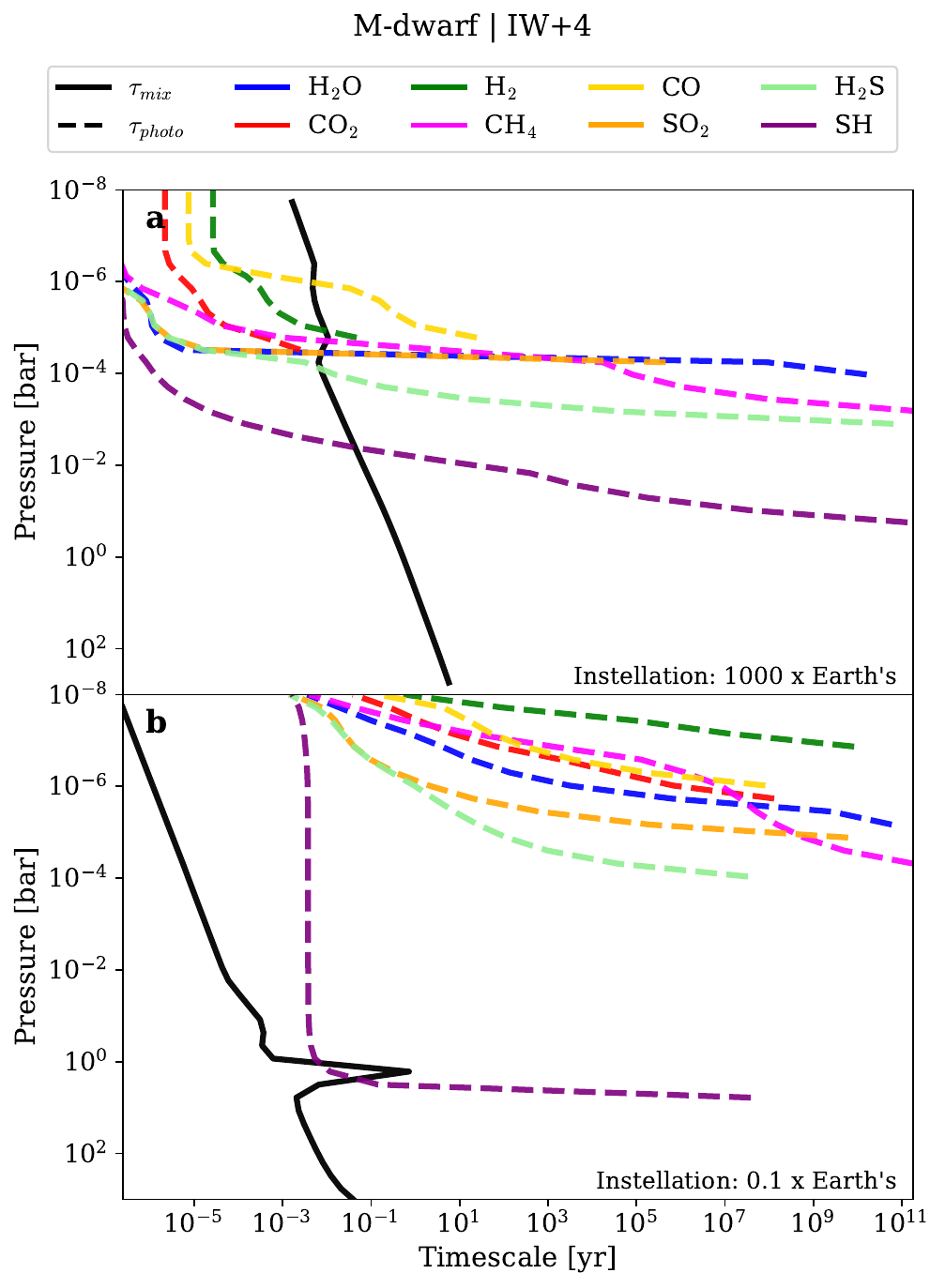}
    \caption{Vertical profiles of the characteristic mixing and photolysis timescales for the selected species in the M-dwarf, IW+4 case, shown for the instellations of (\textbf{a}) $1000\,S_\oplus$ and (\textbf{b}) $0.1\,S_\oplus$. The solid black curve gives the mixing timescale, and the coloured dashed curves indicate the photolysis timescales. Notable species discussed in our work are shown. Where $\tau_{\rm photo}$ falls below $\tau_{\rm mix}$, local chemical processing is faster than vertical transport, and photochemistry has a significant effect on the abundance of each species.}
    \label{fig: timescales}
\end{figure}

The mixing timescale is defined as $\tau_{\rm mix} = H^2/K_{\rm zz}$, where $H$ is the pressure scale height and $K_{\rm zz}$ is the eddy diffusion coefficient \citep{Zhang_2018}. For the most highly irradiated cases (example of IW+4, $1000\,S_\oplus$ in Fig. \ref{fig: timescales} \textbf{a}), $\tau_{\rm mix}$ typically lies in the range $\sim 0.03$--$3$ yr over the $10^{-8}$--$10^{-5}$ bar region. In contrast, for the weakly irradiated cases (example of IW+4, $0.1\,S_\oplus$ in Fig. \ref{fig: timescales} \textbf{b}), it decreases to the order of minutes, implying that vertical transport can dominate the upper-atmosphere composition in those cases. 

As described in Sect. \ref{subsec: top of atm}, $\mathrm{H_2O}$, $\mathrm{CO_2}$, $\mathrm{H_2}$, and $\mathrm{H_2S}$, experience direct photolysis, $\mathrm{H_2O + h\nu \rightarrow OH + H}$, $\mathrm{CO_2 + h\nu \rightarrow CO + O}$, $\mathrm{H_2S + h\nu \rightarrow SH + H}$, with $\mathrm{H_2}$ undergoing indirect chemical destruction through radical reactions such as $\mathrm{H_2 + OH \rightarrow H_2O + H}$. The photolysis timescale is $\tau_{{\rm photo},\rm i} = 1/J_{\rm i}$, where $J_{\rm i}$ is the total photolysis rate of species i, as calculated by \texttt{VULCAN}. Over the $10^{-8}$--$10^{-5}$ bar region, $\tau_{\rm photo}$ increases strongly with decreasing instellation, but increases to infinity for all species before 1 bar, as can be seen from Fig. \ref{fig: timescales} \textbf{a} to \textbf{b}). At $1000\,S_\oplus$ in the M-dwarf cases (Fig. \ref{fig: timescales} \textbf{a}), the median $\tau_{\rm photo}$ value is  $10^{-4}\,\mathrm{yr}$ for $\mathrm{H_2}$ and CO, $10^{-4.5}\,\mathrm{yr}$ for $\mathrm{CO_2}$ and $\mathrm{CH_4}$, $10^{-5}\,\mathrm{yr}$ for $\mathrm{H_2O}$ and $\mathrm{SO_2}$, $10^{-5.5}\,\mathrm{yr}$ for $\mathrm{H_2S}$, and $10^{-6}\,\mathrm{yr}$ for $\mathrm{SH}$. Comparing these values with $\tau_{\rm mix}$, which is typically $10^{-2}$ yr at $1000\,S_\oplus$, shows that photochemical processing dominates the upper atmospheres of the highest-instellation cases. At $0.1\,S_\oplus$ (Fig. \ref{fig: timescales} \textbf{b}), the mixing timescale falls to $10^{-5}$ yr, which is very small compared to the average photolysis timescale of 100 years over all the species. This shows that at low instellation, vertical mixing can more effectively preserve the compositional signature of the deeper outgassed reservoir. As mentioned in Sect. \ref{subsec: overall}, the outgassing timescales lie between $2\times 10^5$ and $10^7$ yr, which is well above the timescales for mixing and photolysis, illustrating the importance of disequilibrium reprocessing in rocky planet magma ocean atmospheres.

The species shown in Fig. \ref{fig: timescales} illustrate that the dominant controlling process can also differ among molecules. Stable major species such as $\mathrm{H_2}$ and $\mathrm{CO_2}$ remain comparatively long-lived ($\sim 10^{-4}$ yr in the most strongly irradiated case, $\sim 10^{3}$ yr in the least irradiated case over $10^{-5}-10^{-8}$ bar), whereas reduced carbon and sulfur intermediates such as $\mathrm{CH_4}$, $\mathrm{CO}$, $\mathrm{H_2S}$, and $\mathrm{SH}$ can transition more abruptly between transport-dominated and chemistry-dominated behaviour ($\sim 10^{-6}$ yr in the most strongly irradiated case, $\sim 10^{1}$ yr in the least irradiated case over $10^{-5}$--$10^{-8}$ bar). This change in their photolysis timescale is most prominent for SH, an intermediate species in the sulfur network (dashed purple lines in Fig. \ref{fig: timescales}), which maintains an average value of $\sim 10^{-2}$ yr as deep as 1 bar. These differences help explain why sulfur chemistry is an important process in the simulated atmospheres presented in this work and their rapid response to irradiation.

\section{Benzene formation mechanism}\label{Asec: benzene}
An important component of our analysis is the use of the equilibrium results as the initial conditions to the atmospheric composition when exploring disequilibrium effects. \texttt{FastChem} is utilised in \texttt{VULCAN} as the initial step before the post-processing applies mixing and photochemistry effects (as explained in Sect. \ref{sec: atmospheric chemistry}). The species recognised in \texttt{VULCAN}'s version of \texttt{FastChem} include benzene, $\mathrm{C_6H_6}$, and a formation pathway through propargyl, $\mathrm{C_3H_3}$, is implemented with the goal of capturing the presence of haze precursors. The formation of benzene is based on the HACA (Hydrogen-Abstraction/Carbon-Addition) mechanism  \citep{Tsai_2021}.

\begin{figure}[t!]
    \centering
    \includegraphics[width=\columnwidth]{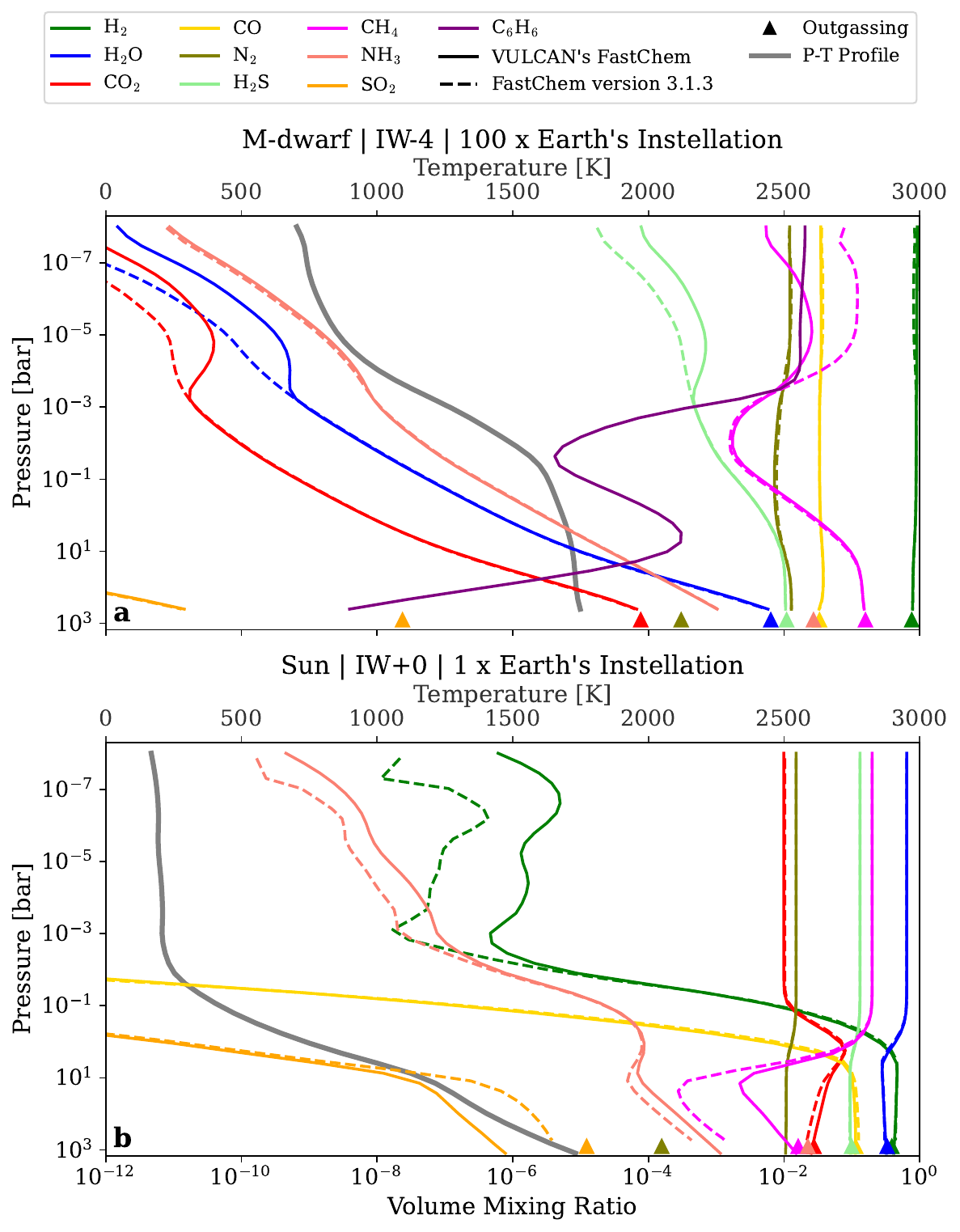}
    \caption{Equilibrium chemistry results for two simulation cases showing the most significant deviation. The results from \texttt{FastChem} version 3.1.3 are shown with the dashed lines, and \texttt{VULCAN}'s version of \texttt{FastChem} is shown with the solid lines, which includes additional hydrocarbons such as $\mathrm{C_6H_6}$ (indicated in purple). The pressure-temperature profile is shown in grey, and the outgassing abundances at the bottom layer is shown with a triangular marker.}
    \label{fig:fastchem_comparison}
\end{figure}

Figure \ref{fig:fastchem_comparison} compares the results with \texttt{FastChem} version 3.1.3 \citep{Kitzmann_2023} for two cases that show notable deviations: IW-4 with $100\,S_\oplus$, and IW+0 equal to Earth's instellation flux. The former (Fig. \ref{fig:fastchem_comparison} \textbf{a}) shows a significant presence of benzene in the upper atmosphere, paired with enhancement of $\mathrm{H_2O}$ and $\mathrm{CO_2}$, and depletion of $\mathrm{CH_4}$. The benzene mechanism introduces an additional carbon sink that redistributes elemental abundances, especially in highly reduced atmospheres (IW-4), where carbon is initially stored in $\mathrm{CH_4}$. As a result, $\mathrm{CH_4}$ is depleted as it is converted into more complex hydrocarbons, while $\mathrm{C_6H_6}$ becomes abundant in the upper atmosphere. This redistribution also impacts other species, such as $\mathrm{H_2S}$ due to competition for hydrogen, while $\mathrm{H_2O}$ and $\mathrm{CO_2}$ increase as oxygen is re-partitioned when less carbon is available in simple molecules. The effect is most pronounced at low pressures, where conditions favour the formation of larger hydrocarbons. Their volume mixing ratios change by up to an order of magnitude at $10^{-5}$ bar. 

A subtle transition to the opposite regime of importance can be observed for the latter case (Fig. \ref{fig:fastchem_comparison} \textbf{b}), where $\mathrm{H_2}$ and $\mathrm{NH_3}$ get enhanced, since a transition to more oxidised cases disfavours reduced hydrocarbons. Therefore, more H is free to bind with $\mathrm{H_2}$ and $\mathrm{NH_3}$. Overall, this demonstrates that the inclusion of complex chemical pathways can modify the equilibrium composition itself by about an order of magnitude, mostly for reduced cases, although it does not qualitatively change the dominance of bulk species or the results presented in this work.

\section{Effects of condensation}\label{Asec: condensation}

The simulations run in our work included Earth-sized rocky planets at low instellation fluxes, specifically equal to Earth's or 10 times less. As outlined in Sect. \ref{subsec: overall}, all the simulations under $100\,S_\oplus$ reached convergence when the mantle solidified. In Fig. \ref{fig:pt_profiles}, the temperature of the 1.0 and $0.1\,S_\oplus$ cases dropped to 150\,K at the top of the atmosphere, whereas the surface reached approximately 1500\,K. These values illustrate the effects of condensation, which was not implemented in our simulations and post-processed atmospheric characterisations.

\begin{figure*}[t]
    \centering
    \includegraphics[width=\textwidth]{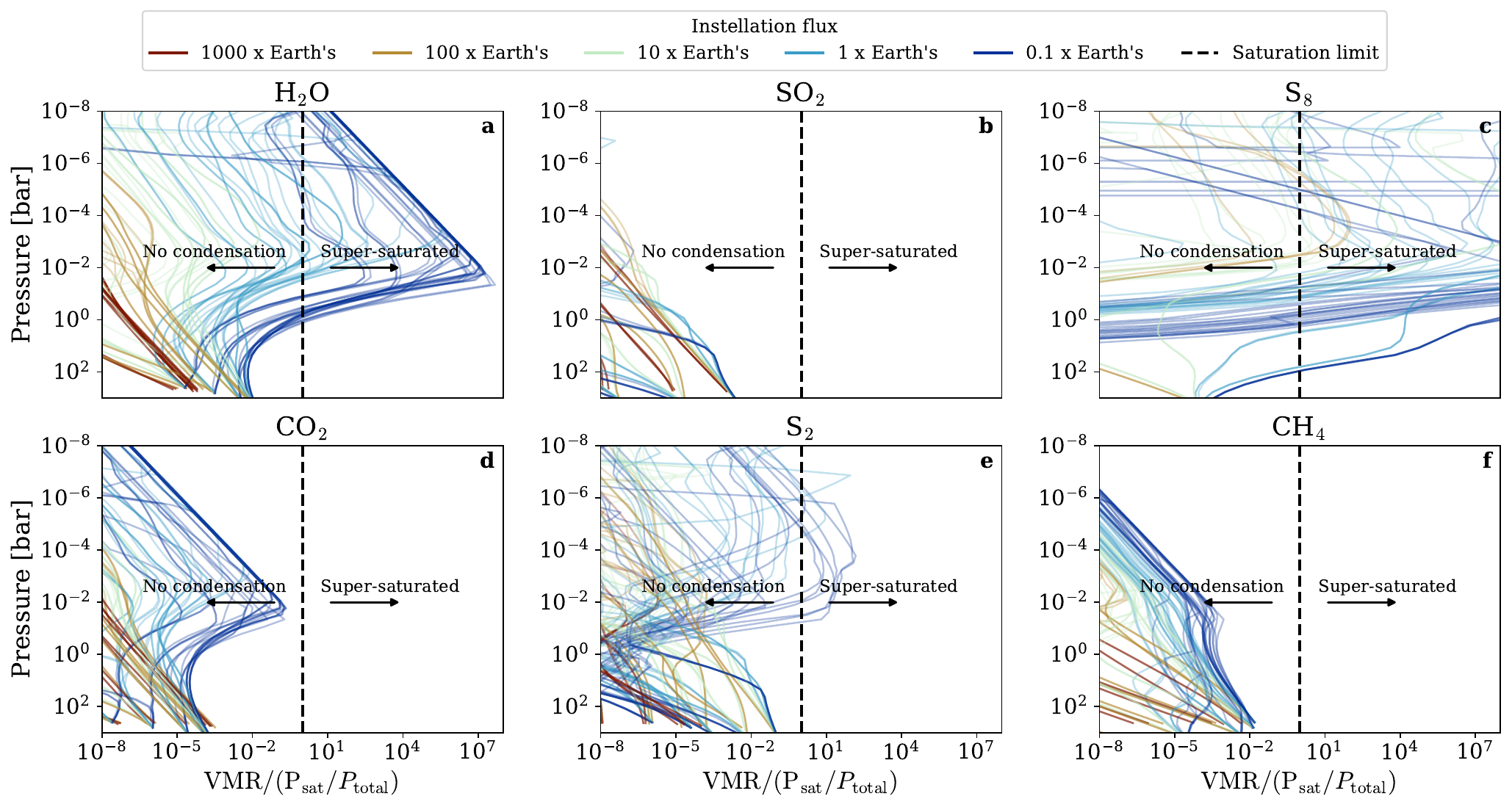}
    \caption{Supersaturation ratio of $\mathrm{H_2O, SO_2, S_8, CO_2, S_2}$, and $\mathrm{CH_4}$ over every atmospheric layer, for all simulations and all atmospheric characterisations considered in our work, colour-coded by the instellation flux.}
    \label{fig:condensation}
\end{figure*}

Figure \ref{fig:condensation} shows the supersaturation ratio, which is defined as the volume mixing ratio over the saturation pressure divided by the total pressure $P_{\rm sat}/P_{\rm tot}$, of 6 species ($\mathrm{H_2O, SO_2, S_8, CO_2, S_2}$ and $\mathrm{CH_4}$) that could condense in the atmospheres explored in our work, for all simulations and all atmospheric characterisations, colour-coded by the instellation flux. The saturation pressure is calculated by integrating the Clausius-Clapeyron relation in the ideal-gas regime with constant enthalpy of vaporisation, and replacing reference parameters for each species as taken from NIST.

$\mathrm{SO_2, CO_2}$, and $\mathrm{CH_4}$ (Fig. \ref{fig:condensation} \textbf{b, d}, and $\textbf{f}$, respectively) stay well below the supersaturation limit over the entirety of the atmosphere, which is defined when the supersaturation ratio becomes equal to 1, indicating that they do not condense for any of the atmospheres. $\mathrm{H_2O}$ (Fig. \ref{fig:condensation} \textbf{a}) becomes supersaturated for cases with 1 and $0.1\,S_\oplus$, crossing the limit at 1 bar, peaking at $10^{-2}$ bar. Similarly, $\mathrm{S_8}$ (Fig. \ref{fig:condensation} \textbf{c}) becomes supersaturated for the lowest instellation flux cases, but includes some cases at $10\times$, too. Finally, $\mathrm{S_2}$ (Fig. \ref{fig:condensation} \textbf{e}) remains mostly close to the limit, with a few cases reaching a supersaturation ratio of $10^2$ between 1 and $10^{-6}$ bar. Figure \ref{fig:condensation} does not show a significant separation in cases when considering the mantle redox state, star type, or atmospheric characterisation, highlighting that condensation is strongly connected to the overall received radiation rather than specific chemistry in our work.

It is indeed expected that the cases with 1 and $0.1\,S_\oplus$ show condensation. Condensation and cold trapping are important processes when considering the evolution of rocky planets in the habitable zone, and therefore important to the photochemical calculations. As shown in Fig. \ref{fig:condensation}, condensation does not happen for all species, but can reach significant amounts for water. Condensation would remove a significant amount of water from the upper atmosphere, resulting in less short-wave absorption and lower temperatures. Therefore, our results are overestimating the upper atmosphere temperature of the low-instellation magma oceans.

In this work, we model the magma ocean phase of rocky planets, with a post-processed chemistry calculation on the outgassed atmosphere to identify potential deviations from equilibrium due to photochemistry. As discussed in Sect. \ref{Asec: timescales}, the atmospheric compositions of very low instellation flux cases are dominated by vertical transport instead of photochemistry. However, the results presented in this work do not explicitly demonstrate the effects of photochemistry on low-instellation magma oceans together with condensation. Coupling condensation and chemistry is tricky and requires strongly coupled microphysics. We leave these questions to be explored in future work. Overall, most cases remain dry, and the results presented in our work are largely robust, at least for cases with $\geq 10\,S_\oplus$.

\end{appendix}

\end{document}